\documentclass{article}
\usepackage{fullpage}
\usepackage{graphicx}
\usepackage{subeqnarray}
\usepackage{amsmath}
\numberwithin{equation}{section}
\usepackage{appendix}
\usepackage{bm}

\title{A model for the electric field-driven deformation of a drop or vesicle in strong electrolyte solutions}

\author{Manman Ma \\
School of Mathematical Sciences, Tongji University, Shanghai 200092, China
\and
Michael R. Booty \& Michael Siegel \\
Department of Mathematical Sciences and Center for Applied Mathematics and Statistics, \\
New Jersey Institute of Technology, 
Newark, New Jersey 07102, USA}

\date{28 July 2021}

\begin{document}

\bibliographystyle{unsrt}

\maketitle

\begin{abstract}
A model is constructed to describe the arbitrary deformation of a drop or vesicle that contains and is embedded in an electrolyte solution, where the deformation is caused 
by an applied electric field.  The applied field produces an electrokinetic flow or induced charge electro-osmosis.  The model is based on the coupled Poisson-Nernst-Planck and 
Stokes equations.  These are reduced or simplified by forming the limit of strong electrolytes, for which ion densities are relatively large, together with the limit of thin Debye layers. 
Debye layers of opposite polarity form on either side of the drop interface or vesicle membrane, together forming an electrical double layer.  

Two formulations of the model are given.  One utilizes an integral equation for the velocity field on the interface or membrane surface together with a pair of integral equations 
for the electrostatic potential on the outer faces of the double layer.  The other utilizes a form of the stress-balance boundary condition that incorporates the double layer structure 
into relations between the dependent variables on the layer's outer faces.  This constitutes an interfacial boundary condition that drives an otherwise unforced Stokes flow outside 
the double layer.  For both formulations relations derived from the transport of ions in each Debye layer give additional boundary conditions for the potential and ion concentrations 
outside the double layer.    

\end{abstract}


\section{Introduction} 
\label{intro}

The purpose of this study is to construct a model based on the Poisson-Nernst-Planck and Stokes equations in the thin Debye layer limit for the deformation of an inclusion that is 
embedded in an exterior medium, where both the inclusion and exterior medium are electrolyte solutions.  The deformation is driven by an imposed electric field and the solutions 
are strong electrolytes.  Two examples are considered side by side.  In one example, the electrolyte solutions and their solvents are immiscible, so that the inclusion is a drop, 
and the interface between the two fluids is assumed to be completely impervious to the passage of ions, so that it is referred to as ideally polarizable.  This is perhaps the simplest 
model for the interface between two immiscible electrolyte solutions (ITIES) of electrochemistry.  In the other example, the same electrolyte solution occupies both the interior 
and exterior phases, which are separated by an interfacial membrane, such as an inextensible lipid bilayer or a vesicle membrane.  The membrane significantly impedes the 
passage of both solvent and ions but still allows a small flux of each species to pass separately between the two phases.

At each point of the interface the Debye layers on its opposite sides have opposite polarity, and together the pair are referred to as an electrical double layer.  
Both the Debye layer and double layer are fundamental concepts in electrochemistry and colloid science, and have been been studied theoretically since the work by Helmholtz 
of 1853 \cite{Helmholtz1853}, 
often in the context of an electrically charged solid in contact with an electrolyte solution.  The underlying structure used here for the distribution of 
ions and the electrostatic potential within each Debye layer is the Gouy-Chapman model, of Gouy \cite{Gouy} and Chapman \cite{Chapman}, in which diffusion of ions is 
balanced by an electrostatic Coulomb force acting on their electrical charge at the continuum level.  Although the model has been developed since to include additional 
effects such as the non-zero size of ions, the influence of the solvation shell around an ion, and other realistic effects, these developments all retain the basic constituents of the 
Gouy-Chapman model at their core.  The texts by Russel {\it et al.} \cite{Russel_etal} and Hunter \cite{Hunter} give a review of the general theory of the Debye and double layer and 
describe related experiments.  

The mechanism for flow and deformation is as follows.  The electric field $\boldsymbol{E}$ in the medium causes ions to move or migrate under the action of an 
electrostatic Coulomb force, with the (positively charged) cations moving in the direction of $\boldsymbol{E}$ and the (negatively charged) anions moving in the direction 
opposite to $\boldsymbol{E}$.  Both ion species are advected with the underlying fluid velocity modified by a molecular drift velocity due to the combined effects of a 
concentration dependent diffusive flux and the Coulomb force induced electromigration.  Away from boundaries and interfaces that impede their motion ions are present in 
number densities according to their valence that maintain zero net charge or electroneutrality.  For simplicity we consider binary symmetric electrolytes with valence 1 ions, 
so that these number densities are equal.  

However, near a boundary or interface that stops or impedes the movement of ions in the normal direction, ion densities change and charge separation occurs.  
Positive charge develops when $\boldsymbol{E}$ has a normal component directed from the medium to the interface and negative charge develops when 
$\boldsymbol{E}$ has a normal component in the opposite direction, that is, directed from the interface into the medium.  An electrical double layer forms, with Debye 
layers of opposite polarity on either side of the interface.  

The presence of separated or induced charge in a net electric field causes a force to be exerted on the fluid locally.  The force has a normal component that can deform 
the interface and a tangential component that induces fluid flow along the interface with accompanying shear stress that can also cause interfacial motion and deformation.  
This is an example, for a deformable interface, of induced charge electrokinetic flow or induced 
charge electro-osmosis (ICEO).  It is described in the above-mentioned texts by Russel {\it et al.} \cite{Russel_etal} and Hunter \cite{Hunter}, and more recent reviews of the 
theory and experiments of ICEO have been given by, for example, Squires \& Bazant \cite{Squires_Bazant_04,Bazant_Squires_10}.  Girault \cite{Girault_2010} reviews 
the electrochemistry of liquid-liquid interfaces (i.e., ITIES) and Reymond {\it et al.} \cite{Reymond_&co}  discuss current and potential applications at their time of writing. 

Much of the recent interest in electrokinetic flow and the related area of electrohydrodynamics (EHD) is motivated by their applications in microfluidics;   Vlahovska
\cite{Vlahovska2019} gives a comprehensive review that addresses fundamental aspects of many studies in EHD.  Central to electrohydrodynamics is the 
leaky dielectric model of Melcher \& Taylor \cite{Melcher_Taylor_1969}, and a contrast between electrohydrodynamics and electrokinetics is described in the introduction 
of the review by Saville \cite{Saville_1997}.  Namely, in EHD electric charge is situated at or on the interface between media of different electric permittivity that are weak 
conductors of charge due to a relatively low concentration of ions.  Large electric field strengths must be applied to induce a flow in these weak electrolytes.  
In electrokinetics, charge-carrying ions are plentiful and present in a diffuse charge cloud near a surface.  An applied potential of the order of volts per centimeter is sufficient 
to induce flow in these strong electrolytes.  

The focus of the study by Saville \cite{Saville_1997} is a derivation of the leaky dielectric model based on the governing equations of electrokinetics, the 
Poisson-Nernst-Planck equations, that is applicable to weak electrolytes.  In this sense, he explains there, the two topics merge.  Further, this had been anticipated   
by Taylor \cite{Taylor_1966} in the concluding remarks of his study on the circulation induced in a drop by an electric field, where he comments that the predictions 
of the electrohydrodynamic model `may be expected to be realistic even if the charge is not exactly situated at the interface, provided its distance from the interface 
is small compared with the linear scale of the situation'.

Since Savile \cite{Saville_1997} two more recent studies on the derivation of the leaky dielectric, electrohydrodynamic model as a limiting form for weak electrolytes of the 
Poisson-Nernst-Planck equations of electrokinetics have been made by Schnitzer \& Yariv \cite{SchnitzerYariv} and by Mori \& Young \cite{MoriYoung}.  

Pascall \& Squires \cite{PascallSquires2011} give a detailed study of electrokinetic phenomena at liquid-liquid interfaces.  Their main focus is the dependence of the free-stream 
or slip velocity of an electrolyte solution, far from a double layer, on characteristic properties of the underlying media.  In particular, given an imposed tangential electric field, 
there is an increase in the free-stream velocity by a factor of $d/\lambda_{*}\gg1$ if a liquid film of thickness $d$ is introduced between the electrolyte solution and 
an underlying solid metal substrate, where $\lambda_{*}$ is the thickness of the Debye layer.  This occurs when the liquid film is either a perfect conductor or a dielectric 
permeated by an electric field.  Their study compares and successfully reconciles the results and predictions of earlier studies, and establishes similar results for the 
electrophoresis of spherical liquid drops.  The geometry is fixed, either planar or spherical, and the context is that of a steady state.  

The present study is organized as follows.   In \S \ref{sec:formulation} the governing field equations and boundary conditions, including the conditions at a drop and vesicle 
interface are stated, together with the initial conditions and the far-field conditions for an imposed electric field.  The field equations consist of the Poisson-Nernst-Planck 
equations coupled to the equations of zero-Reynolds number or Stokes flow.  Nondimensional scales are introduced and the nondimensional form of the system is 
given in \S \ref{sec:nondimall}.  The Nernst-Planck equations for the ion concentrations include rate terms that express the rate of dissociation of salt or 
electrolyte into ions and the recombination or association of ions into salt.  In \S \ref{sec:SEL} the limit of a strong electrolyte is formed, for which all or nearly 
all of the salt is present in its dissolved or disassociated form as ions.  The analysis then proceeds with the ion concentration equations in an effectively rate-free form.  

In the thin Debye layer limit, a representative Debye layer thickness $\lambda_{*}$ is much less that the linear dimension $a$ of the inclusion, so that their ratio 
$\epsilon=\lambda_{*}/a$  is small, that is, $0<\epsilon \ll 1$.  The $\epsilon\rightarrow 0$ limit is taken up first in \S \ref{sec:outer}, where the field equations that 
apply in the outer regions, away from the Debye layers, are given at leading order in an $\epsilon$-expansion.  These equations are devoid of source or inhomogeneous 
forcing terms.  

Section \ref{sec:layers1} concerns the structure of the inner regions or Debye layers.  A surface-fitted intrinsic coordinate system is introduced 
(\S \ref{sec:intrinsic}).  With this, the Gouy-Chapman solution for the ion concentrations and electrostatic potential within the layers is constructed, at leading order, in a form 
that is parameterized by variables at the outer edges of the Debye layers (\S \ref{sec:pnp0}).  Expressions for the pressure and fluid velocity in terms of the local 
or excess potential within the layers are given in \S \ref{sec:hydro0}, with a similar parameterization,.  

For two immiscible electrolyte solutions, a drop possesses a charged double layer even when in equilibrium, and the accompanying jump in potential across the double 
layer is referred to as an inner or Galvani potential \cite{Hunter}.  This potential, via the Debye layer $\zeta$-potentials, is expressed in terms of ion partition coefficients
in the reduced or small-$\epsilon$ limit in \S \ref{sec:Galvani}.  In \S \ref{sec:hydro0a} reduced expressions are given for the small but non-zero trans-membrane 
ion flux and osmotic solvent flux across a vesicle membrane.  

To form a closed, reduced asymptotic or macroscale model it is necessary to consider the transport of ions within the Debye layers, which is the subject of 
\S \ref{sec:ion_transport}.  Sections \ref{sec:ITdrop} and \ref{sec:ITvesicle} give the transport relations that are specific to a drop and to a vesicle, and their use as 
interfacial boundary conditions is taken up in \S \ref{sec:outer_c} 

In \S \ref{sec:uint} 
a Fredholm second kind integral equation is derived that gives the fluid velocity on the interface in terms of a net interfacial traction and an integral term that depends on 
the electrostatic energy density contained in the double layer and which has a Stokes-dipole kernel.  Some details of the construction are put in Appendix \ref{sec:appA}. 
Section \ref{sec:phi_int} recalls the integral equation for the electrostatic potential.  

A formulation via integral equations is convenient for simulating large-amplitude deformations.  For small-amplitude deformation the stress-balance boundary condition at 
the interface can be used instead, and this, expressed in terms of quantities at the outer edges of the Debye layers and known surface data, is given in Appendix 
\ref{sec:appB} for a drop and for a vesicle.  The location in the text of relations that are needed to form the model is summarized in \S \ref{sec:summary}, and 
concluding remarks are made in \S \ref{sec:conclusion}.

\section{Formulation: governing equations and boundary conditions} 
\label{sec:formulation} 

The formulation begins with the Nernst-Planck equations for a dilute electrolyte solution coupled to the Stokes equation for an incompressible fluid.  
For a symmetric $1:1$ binary electrolyte, with one positive ion species (or cation) of charge $e$ and one negative ion species (or anion) of the same valence, 
the Nernst-Planck equations for ion  conservation are  
\begin{subeqnarray}
\partial_t c^{(\pm)} + \nabla\cdot (c^{(\pm)}\boldsymbol{v}^{(\pm)})={\cal R} \, , \hspace{10mm}  \slabel{NPdima} \\[3pt] 
\boldsymbol{v}^{(\pm)} - \boldsymbol{u} = \mp b^{(\pm)}e\nabla\phi - D^{(\pm)}\nabla \ln c^{(\pm)} ,  \slabel{NPdimb}
\label{NPdim}
\end{subeqnarray}
with conservation of the combined salt given by 
\begin{subeqnarray}
\partial_t s + \nabla\cdot (s\boldsymbol{v})= - {\cal R} \, , \hspace{8mm} \slabel{sdima}  \\[3pt] 
\boldsymbol{v} - \boldsymbol{u} =  - D_{s}\nabla \ln s . \hspace{10mm}  \slabel{sdimb}
\label{sdim}
\end{subeqnarray}
Quantities specific to an ion species are given a $(\pm)$-superscript to denote the ion charge, $(+)$ for the cation and $(-)$ for the anion, and $c$ denotes 
the ion species concentration.  The Eulerian velocity of a species is denoted by $\boldsymbol{v}$ and the mass averaged fluid velocity is $\boldsymbol{u}$.  

Equation (\ref{NPdimb}) gives the drift velocity of ions through the fluid.  This is due to (i) the Coulomb force on ion charges in the electric field, 
which induces electromigration, with ion mobility $b^{(\pm)}$ and electrostatic potential $\phi$, and (ii) the diffusion of ions due to variations in their 
concentration, with ion diffusivity given by the Einstein-Smoluchowski relation $D^{(\pm)}=b^{(\pm)}k_{B}T$.  The drift velocity is also the product of the ion 
mobility and the net thermodynamic force acting on the ions, which is minus the gradient of their electrochemical potential, and (\ref{NPdimb}) is referred 
to as the Nernst-Planck relation \cite{Russel_etal}.  It can be used to recast the expression for the molecular ion flux, which is 
\begin{equation}
\boldsymbol{j}^{(\pm)}=c^{(\pm)}(\boldsymbol{v}^{(\pm)} - \boldsymbol{u}) \, .
\label{jdef0}
\end{equation} 
The concentration of the 
electrically neutral combined salt is denoted by $s$, with diffusivity $D_{s}$.  The rate term ${\cal R}$ is given by 
\begin{equation}
{\cal R}= k_{d}s -k_{a} c^{(+)}c^{(-)} \, ,
\label{R}
\end{equation}
where $k_{d}$ is the rate of dissociation of the salt into ions and $k_{a}$ is the rate of association of ions into salt.  

An $i$-subscript is used to denote separate phases, with $i=1$ for the dispersed or interior phase domain $\Omega_{1}$ and $i=2$ for the continuous or 
exterior phase domain $\Omega_{2}$.  Later, an $i$-subscript will also be applied to material parameters and ambient species concentrations, which can be 
different in the two phases but are constant within each phase.  However, it will be omitted initially when the distinction does not need to be made.   

The electric displacement $\boldsymbol{D}$, electric field $\boldsymbol{E}$, and volumetric charge density $\rho$ satisfy the relations 
 \begin{equation}
\nabla\cdot \boldsymbol{D}=\rho, \ \ \ \boldsymbol{D}=\varepsilon \boldsymbol{E}, \ \ \ \boldsymbol{E}=-\nabla \phi, \ \ \ 
\rho=e(c^{(+)}-c^{(-)}) .
\label{Efield}
\end{equation}
These are respectively Gauss's law, a linear constitutive relation with relative electric permittivity or dielectric constant $\varepsilon$, 
the relation between the electric field and the electrostatic potential, and the charge density in terms of the ion densities.  Together they give 
the Poisson equation in the form 
\begin{equation}
\nabla^{2}\phi= -\frac{e}{\varepsilon}(c^{(+)}-c^{(-)}).
\label{Pdim}
\end{equation} 

In the Stokes flow or zero Reynolds number limit for an incompressible electrolyte, 
\begin{equation}
\nabla \cdot \boldsymbol{u}=0,\ \ \mbox{and} \ \ -\nabla p +\mu\nabla^{2}\boldsymbol{u} + \rho \boldsymbol{E} = 0 , 
\label{Stokesdim}
\end{equation}
where $p$ is the pressure.  
The Coulomb force term $\rho \boldsymbol{E}$ in the Stokes equation provides the coupling from the electrostatic field or Nernst-Planck and 
Poisson equations to the fluid field, and can be expressed in various equivalent ways by use of (\ref{Efield}).  

A sketch of the Debye layer pair near the interface, which introduces some of the notation we use, is given in Figure \ref{fig:setup}.

\subsection{Boundary Conditions for a Drop} 

In the boundary conditions adopted for a drop, the interface has no electrical capacitance and no monopole surface charge.  It is also impervious or impenetrable 
to ions, and is therefore referred to as ideally polarizable.  This gives the interfacial boundary conditions
\begin{equation}
[\phi]=0, \ \ \ [\boldsymbol{D}\cdot\boldsymbol{n}]=0, \ \ \mbox{and}  \ \ \boldsymbol{j}^{(\pm)} \cdot\boldsymbol{n}=0 \ \ \  \mbox{on} \ S ,
\label{BC1dim}
\end{equation}
respectively.  Here $S$ denotes the sharp interface between the two phases, with outward unit normal $\boldsymbol{n}$, and 
$[\cdot]\equiv [\cdot]^{\partial \Omega_{2}}_{\partial \Omega_{1}}$ denotes the 
jump across $S$, with the convention that it is the limit as $S$ is approached from the exterior domain $\Omega_{2}$ minus the limit 
as $S$ is approached from the interior domain $\Omega_{1}$.  

Continuity of velocity of the solvent and the kinematic boundary condition imply that 
\begin{equation}
[\boldsymbol{u}]=0, \ \ \mbox{and} \ \ (\boldsymbol{u}-\frac{\partial\boldsymbol{x}_{s}}{\partial t})\cdot\boldsymbol{n}=0 \ \ \mbox{on} \ S,  
\label{BC2dim}
\end{equation}
where $\boldsymbol{x}_{s}$ is any point on $S$.  The stress-balance boundary condition is that 
\begin{eqnarray}
[(\boldsymbol{T}_{H} + \boldsymbol{T}_{M})\cdot\boldsymbol{n}] = \sigma(\kappa_{1}+\kappa_{2})\boldsymbol{n} 
                   \ \ \mbox{on} \ S, \hspace{20mm} 
\label{sbaldim} \\
\mbox{where} \ \ \boldsymbol{T}_{H}= -p\boldsymbol{I}+2\mu\boldsymbol{e} \ \ \mbox{and} \ \ 
\boldsymbol{T}_{M}= \varepsilon(\nabla \phi \otimes \nabla \phi-\frac{1}{2}|\nabla\phi|^{2}\boldsymbol{I}) .
\label{tensdim}
\end{eqnarray}
Here $\boldsymbol{T}_{H}$ is the stress tensor for a Newtonian fluid with viscosity $\mu$ and 
$(\boldsymbol{e})_{ij}=(\partial_{x_{j}}u_{i}+\partial_{x_{i}}u_{j})/2$ is the symmetric part of the velocity gradient.  $\boldsymbol{T}_{M}$ 
is the part of the Maxwell stress tensor due to the electric field, which has been written in terms of the potential $\phi$, $\sigma$ is a 
constant surface tension, and $\kappa_{1}$ and $\kappa_{2}$ are the principal curvatures of $S$.  With $\boldsymbol{n}$ pointing outward on $S$ 
the principal curvatures are taken to be positive when the curve of intersection of S by a plane containing $\boldsymbol{n}$ and a principal direction is convex 
on the side to which $\boldsymbol{n}$ points, and is negative otherwise.  

Under equilibrium conditions there is typically a difference in the electric potential between distinct phases that occurs at and near the interface between them.  
It is caused by a difference in the affinity for charge carriers of the phases and is referred to as an inner or Galvani potential \cite{Hunter}.  
The boundary conditions  
\begin{equation}
c^{(+)}|_{\partial\Omega_{1}} = l^{(+)} c^{(+)}|_{\partial\Omega_{2}}  , \ \mbox{and} \  c^{(-)}|_{\partial\Omega_{1}} = l^{(-)} c^{(-)}|_{\partial\Omega_{2}}  
   \ \ \mbox{on} \ S \, , 
\label{G1}
\end{equation}
where $l^{(+)}$ and $l^{(-)}$ are constant partition coefficients, imply the presence of  a non-zero Galvani potential when $l^{(+)}\neq l^{(-)}$.  The notation 
$c^{(\pm)}|_{\partial\Omega_{i}}$ denotes evaluation of the ion concentration $c^{(\pm)}$ in the limit as $S$ is approached from the domain 
$\Omega_{i}$ for $i=1,2$.  Boundary conditions of this type have been considered by, for example, Zholkovskij {\it et al.} \cite{Zholkovskij} and by 
Mori \& Young \cite{MoriYoung}.

\subsection{Boundary Conditions for a Cell or Vesicle} 

Mori {\em et al.} \cite{Mori_Liu_Eisenberg} introduce a set of phenomenological boundary conditions for the behavior of a cell or vesicle, where the interior and 
exterior phases are separated by a membrane, and these are summarized here.  The membrane has an initial equilibrium reference configuration $S_{ref}$ 
with a local coordinate system $\boldsymbol{\theta}$ that serves as a Lagrangian or material coordinate.  At subsequent times the membrane, which is treated 
as a sharp interface $S$ between the phases $\Omega_{i}$ that nonetheless has electromechanical structure, is described by 
$\boldsymbol{x}=\boldsymbol{X}(\boldsymbol{\theta}, t)$.

The same notation for interfacial quantities is used for both a drop and a vesicle.  Here we have:  $[\cdot]$ for the jump in a quantity across the membrane, 
$\cdot|_{\partial\Omega_{i}}$ for evaluation as the side $\partial\Omega_{i}$ of the membrane is approached, or simply on $S$ for a quantity that is continuous.

If the membrane is semi-permeable or porous to the aqueous solvent it allows a flow of solvent in the normal direction that occurs by osmosis and is assumed 
to be proportional to the jump in the solvent's partial pressure across the membrane.  Hence  
\begin{equation}
\boldsymbol{u}-\left.\frac{\partial\boldsymbol{X}}{\partial t}\right|_{\boldsymbol{\theta}} = - \, \pi_{m} \, [ p- k_{B}T (c^{(+)}+c^{(-)}) ] \, \boldsymbol{n} 
               \ \ \mbox{on} \ S \, ,
\label{solfluxdim}
\end{equation} 
where $\pi_{m} \ge 0$ is an effective membrane porosity and the partial pressure of the ions is given by the ideal gas law.  

The trans-membrane flux for each ion species is taken to be proportional to the jump in its electrochemical potential across the membrane surface.  
Mori {\it et al.} \cite{Mori_Liu_Eisenberg} give the boundary condition 
\begin{equation}
c^{(\pm)} (\boldsymbol{v}^{(\pm)} - \left.\frac{\partial\boldsymbol{X}}{\partial t}\right|_{\boldsymbol{\theta}}) \cdot \boldsymbol{n} = 
        - g^{(\pm)} [ k_{B}T \ln c^{(\pm)} \pm e \phi ] 
                       \ \ \mbox{on} \ S \, ,
\label{ionfluxdim1} 
\end{equation}
where $g^{(\pm)}\ge 0$ is a gating constant for each ion species, and the velocity on the left hand side, 
$\boldsymbol{v}^{(\pm)}-\partial_{t}\boldsymbol{X}|_{\boldsymbol{\theta}}$, is the velocity of ions relative to the membrane material.  

In the analysis that follows, a boundary condition is needed for the normal component of the molecular ion flux $\boldsymbol{j}^{(\pm)}\cdot\boldsymbol{n}$ 
in the electrolyte solution immediately adjacent to the membrane, where the flux $\boldsymbol{j}^{(\pm)}$ is given by equation (\ref{jdef0}), i.e., 
$\boldsymbol{j}^{(\pm)}=c^{(\pm)}(\boldsymbol{v}^{(\pm)} - \boldsymbol{u})$.  If the normal ion flux $\boldsymbol{j}^{(\pm)}\cdot\boldsymbol{n}$ 
in the electrolyte and the normal ion flux across the membrane given by (\ref{ionfluxdim1}) are not equal then charge is not conserved at the membrane 
boundaries $\partial\Omega_{i}$ but accumulates there.  We note that this occurs if both (\ref{solfluxdim}) and (\ref{ionfluxdim1}) are applied to a membrane 
that is semi-permeable to solvent, i.e., when $\pi_{m} > 0$.  This is verified by subtracting equation (\ref{solfluxdim}) multiplied by the 
ion concentration $c^{(\pm)}$ from (\ref{ionfluxdim1}) to form the electrolyte ion flux $\boldsymbol{j}^{(\pm)}\cdot\boldsymbol{n}$, and noting that 
the concentration of each ion species is not continuous but has a jump across the membrane.  

This difficulty is easily resolved by modifying the trans-membrane flux boundary condition (\ref{ionfluxdim1}) to read 
\begin{equation}
c^{(\pm)} (\boldsymbol{v}^{(\pm)} - \boldsymbol{u}) \cdot \boldsymbol{n} = 
        - g^{(\pm)} [ k_{B}T \ln c^{(\pm)} \pm e \phi ] 
         \ \ \mbox{on} \ S \, ,
\label{ionfluxdim2} 
\end{equation}
which conserves charge across the membrane and its boundaries and holds in general, that is for all $\pi_{m} \ge 0$.   When the membrane is impermeable 
to solvent (i.e., $\pi_{m}=0$) the boundary condition (\ref{solfluxdim}) reduces to no-slip between the electrolyte and membrane surface, and (\ref{ionfluxdim1}) 
and (\ref{ionfluxdim2}) are identical, but when the membrane is semi-permeable to solvent (i.e., $\pi_{m} > 0$) the solvent and ion species cross the membrane 
at different rates via distinct pores or channels, and (\ref{ionfluxdim2}) must be applied instead.  

\begin{figure}
\begin{center}
\includegraphics[scale=0.27]{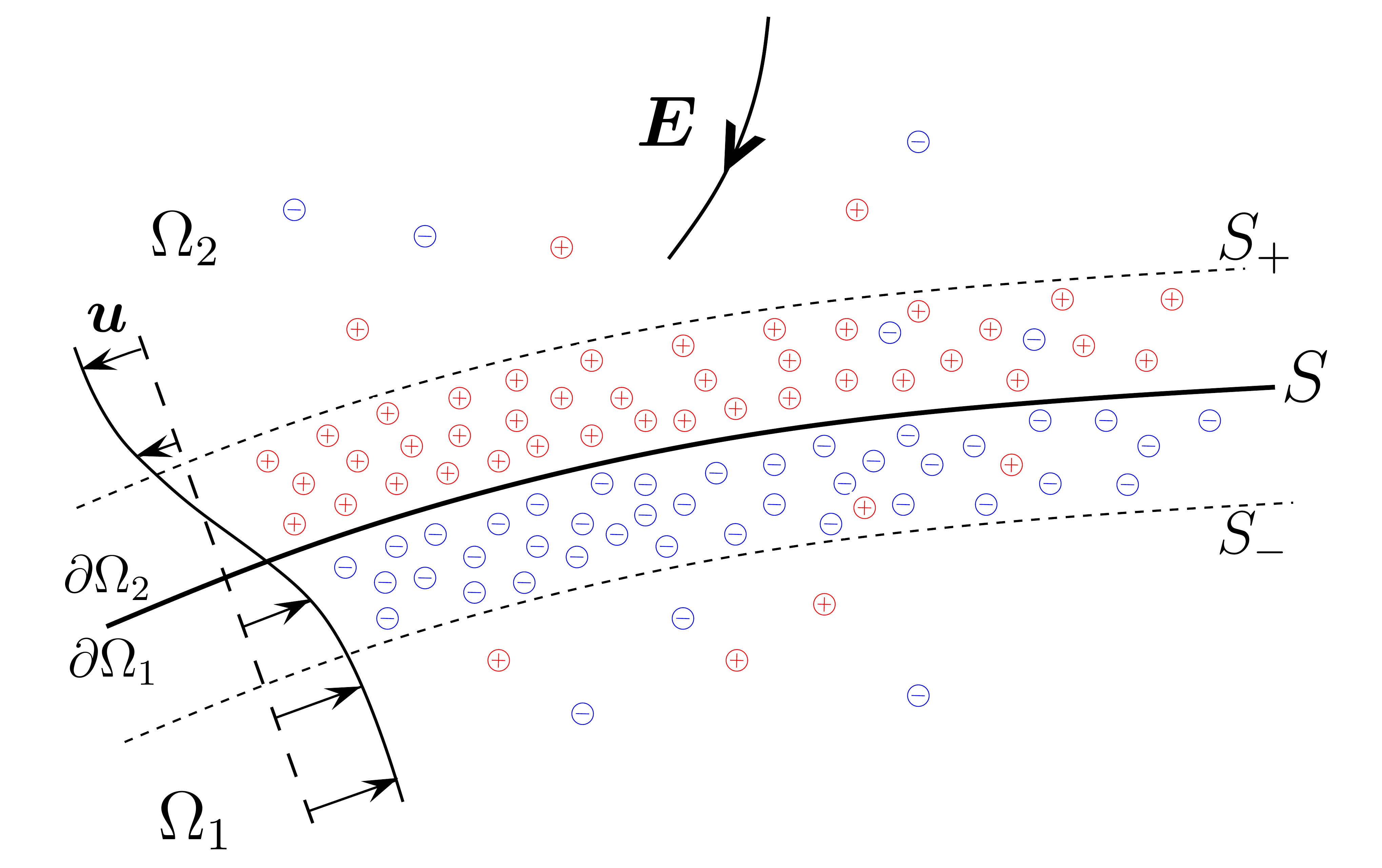} 
\end{center}
\caption{An interior phase $\Omega_{1}$ with boundary $\partial\Omega_{1}$ and an exterior phase $\Omega_{2}$ with boundary $\partial\Omega_{2}$ 
are separated by the interface $S$ of a drop or a vesicle membrane that is either impervious or nearly impervious to the passage of ions.  An electric field with 
strength $\boldsymbol{E}$ induces a flow of ions that charges up a pair of Debye layers of opposite polarity.  The outer edges of the Debye layers are denoted 
by $S_{+}$ in the exterior phase and $S_{-}$ in the interior phase, and the induced flow velocity of the host solvent is denoted by $\boldsymbol{u}$.
}  
\label{fig:setup}
\end{figure}

The membrane has electrical capacitance because it can maintain a jump in the potential across its outer faces, between which the potential varies linearly with 
normal distance.  The jump in the potential is referred to as the trans-membrane potential and is denoted by $[\phi]$.  The membrane has no net monopole charge, 
in the sense that at distinct points along the normal the surface charge densities on opposite faces have equal magnitude and opposite polarity, cancelling each other.  
The charge on each face is related to the trans-membrane potential by the membrane capacitance per unit area $C_{m}$.  Analogous to the first two boundary conditions 
of (\ref{BC1dim}) for a drop, but modified to express continuity of electric displacement and the capacitance of the membrane, we have the vesicle boundary conditions 
\begin{equation}
\varepsilon \nabla \phi\cdot\boldsymbol{n}|_{\partial\Omega_{1}} = \varepsilon \nabla \phi\cdot\boldsymbol{n}|_{\partial\Omega_{2}} = C_{m} [\phi] \, , \ 
\mbox{where}  \ C_{m} =\frac{\varepsilon_{m}}{d}\, . 
\label{capdim}
\end{equation}
Here $\varepsilon_{m}$ is the membrane permittivity, and $d$ is its thickness.  

Two specific membrane capacitance models of Mori {\it et al.} \cite{Mori_Liu_Eisenberg} are  
\begin{equation}
C_{m} = 
\left\{
\begin{array}{ll}
C_{m}^{0} & (\mbox{a}) \, , \\
C_{m}^{0} \delta A (\boldsymbol{X})  & (\mbox{b}) \, . 
\end{array}
\right.  
\label{capexd}
\end{equation}
These hold if, for example, the bulk material of the membrane is incompressible, and then (\ref{capexd}a) applies when the membrane surface is 
locally inextensible and (\ref{capexd}b) applies when it is locally extensible.  The capacitance per unit area in the initial undeformed or reference 
state is $C_{m}^{0}$, which is constant for a homogeneous membrane.  The ratio of the local surface area in the deformed state to its value in the 
initial reference state following a material point $\boldsymbol{x}=\boldsymbol{X}(\boldsymbol{\theta}, t)$ is denoted by $\delta A (\boldsymbol{X})$.  
Bulk incompressibility implies that the volume of a membrane material element $d(\boldsymbol{X}) \delta A (\boldsymbol{X})$ is conserved 
during deformation, so that for the locally inextensible surface of (\ref{capexd}a) $\delta A (\boldsymbol{X})=1$ with  $d(\boldsymbol{X})$  and 
$C_{m}^{0}$ conserved, while for (\ref{capexd}b) $d(\boldsymbol{X})^{-1} \propto \delta A (\boldsymbol{X})$.

The stress-balance boundary condition is 
\begin{equation}
[(\boldsymbol{T}_{H} + \boldsymbol{T}_{M})\cdot\boldsymbol{n}] = 2\sigma H \boldsymbol{n} - \nabla_{s} \sigma 
              - \kappa_{b} \left( (2H-C_{0})(H^{2}-K+\frac{C_{0}H}{2}) + \nabla_{s}^{2} H  \right) \boldsymbol{n} 
\label{cellsbaldim} \\
\end{equation}
on $S$.  The vesicle membrane has a bending stiffness that is expressed in terms of the bending modulus $\kappa_{b}$, the mean curvature 
$H=(\kappa_{1}+\kappa_{2})/2$, the Gaussian curvature $K=\kappa_{1}\kappa_{2}$, and a spontaneous curvature $C_{0}$.  The spontaneous 
curvature $C_{0}$ is twice the mean curvature of the membrane material in a traction-free equilibrium state, see for example Vlahovska {\it et al.} \cite{Vlahovska2009} 
and Seifert \cite{Seifert1999}, and it is usually set to zero.  Here, $\nabla_{s}$ denotes the surface gradient operator, $\nabla_{s}^{2}$ is the Laplace-Beltrami 
or surface Laplacian operator, and the convention for the sign of the principal curvatures $\kappa_{1}$ and $\kappa_{2}$ is the same as that in 
(\ref{sbaldim}).  

The tension in the membrane surface is denoted by $\sigma$ in (\ref{cellsbaldim}) but, in contrast to its appearance in (\ref{sbaldim}) for a drop, the value 
of $\sigma$ is not a given constant.  For a locally inextensible membrane $\sigma$ is determined by imposing inextensibility as a constraint, namely 
\begin{equation}
\nabla_{s}\cdot\left( \frac{\partial}{\partial t} \boldsymbol{X}_{s}(\boldsymbol{\theta}, t)\right) =0 \, , 
\label{inextdim}
\end{equation}
or equivalently $\delta A (\boldsymbol{X})=1$, for all $t>0$.  Here $\partial_{t}\boldsymbol{X}_{s}$ denotes the tangential projection of the velocity 
$\partial_{t}\boldsymbol{X}$ of a material particle onto $S$. For a locally extensible membrane a stress-strain equation of state must be introduced, see  
for example \cite{BBiesel2011}.  The membrane tension is initially constant, but during deformation it can vary from point to point on $S$, leading to the 
term $\nabla_{s} \sigma$ in (\ref{cellsbaldim}).  

We note that Mori {\it et al.} \cite{Mori_Liu_Eisenberg} consider a membrane bending stress given by the variational derivative of a general energy functional.  The specific 
choice of the functional due to Helfrich \cite{Helfrich1973} has been adopted in many studies of cell and vesicle deformation, and leads to the bending stress term 
on the right hand side of (\ref{cellsbaldim}).   Barth\`{e}s-Biesel \cite{BBiesel2011} gives a concise description of the different mechanical properties of vesicle, cell, and capsule 
membranes.

\subsection{The Far-Field and Initial Conditions} 

The electric field that induces flow and deformation is applied at time $t=0$.  It can be spatially uniform and constant, or more general, so that 
\begin{equation}
\boldsymbol{E}\rightarrow 
\left\{
\begin{array}{l}
E_{\infty}\boldsymbol{e}_{z} \ \ \mbox{for a uniform field} \\
-\nabla \phi_{\infty} \ \ \mbox{in general}
\end{array}
\right. 
\hspace{5mm}  \ \mbox{as} \  |\boldsymbol{x}|  \rightarrow \infty \ \mbox{for} \ t > 0,
\label{Eapplieddim}
\end{equation}
where $\phi_{\infty}$ is a solution of Laplace's equation.  Also 
\begin{equation} 
\boldsymbol{u}(\boldsymbol{x}, t) \rightarrow 0 \ \mbox{and} \ \ c^{(\pm)}(\boldsymbol{x}, t) \rightarrow \ \mbox{constant as} \ |\boldsymbol{x}|\rightarrow \infty \ \mbox{for} \ t > 0\, . 
\label{ics}
\end{equation}

A drop is assumed to have relaxed to a spherical shape for $t<0$, with no applied field and equilibrium initial conditions given by 
\begin{equation}
\boldsymbol{u}(\boldsymbol{x}, 0)=0 \ \ \mbox{and} \ \ c^{(\pm)}(\boldsymbol{x}, 0)= \mbox{piecewise constant for all} \ \boldsymbol{x} \, .
\label{ics}
\end{equation}
However, when there is a nonzero Galvani potential difference between the two phases the $t<0$ equilibrium relations for the potential and ion concentrations 
are revised to accommodate an electrical double layer - this will be considered later, in \S \ref{sec:Galvani}.

\subsection{Nondimensionalization}
\label{sec:nondimall}

\begin{table} 
\rule{165mm}{0.2mm}
  \begin{center}
\def~{\hphantom{0}}
  \begin{tabular}{cccc} 
       variable                 & nondimensionalisation  &   variable & nondimensionalisation \\[3pt]
     $\boldsymbol{x}$    & $a$                                &    $p$   &  $\mu_{*} U_{c}/a$  \\
     $\boldsymbol{u}$    & $U_{c}\equiv D_{*} / \lambda_{*}$ & $\phi$ & $\phi_{*}=k_{B}T/e $ \\
     $t$ &  $a/U_{c}$   & $c^{(\pm)}, s$ & $c_{*}, s_{*}$     \\
  \end{tabular}
  \caption{Variables are made nondimensional by the quantities shown.  See text for explanation.}
  \label{tab:nondim}
  \end{center}
\rule{165mm}{0.2mm}
\end{table}

The quantities used to put the equations in nondimensional form are listed in Table \ref{tab:nondim}.  The length scale $a$ is the initial drop radius, or a 
similar length scale for a vesicle.  The velocity scale $U_{c}=D_{*}/\lambda_{*}$ is the ion-diffusive scale, where $D_{*}$ is a representative ion diffusivity 
and $\lambda_{*}$ is the Debye layer thickness.  The time scale $a/U_{c}$ is the scale on which the Debye layers acquire charge after $t=0$, i.e., the 
charge-up time scale.  The scale for the pressure $p$ is the scale of the Stokes flow regime, and the scale for the potential is the thermal voltage 
$k_{B}T/e$, where $k_{B}$ is the Boltzmann constant and $T$ is the temperature.  

All material and rate parameters in the two phases are made nondimensional by a single representative value.  For example, the ion concentrations are 
made nondimensional by a representative ambient ion concentration $c_{*}$, and the salt concentration is made nondimensional by $s_{*}$.  Similarly, 
$D^{(\pm)}, k_{d}, k_{a}, \varepsilon$, and $\mu$ are made nondimensional by $D_{*}, k_{d*}, k_{a*}, \varepsilon_{*}$, and $\mu_{*}$ respectively. 
To observe equilibrium between the ion and salt concentrations, the relation 
\begin{equation}
k_{d*}s_{*}=k_{a*}c_{*}^{2}
\label{rateeqmdim}
\end{equation}
holds.  The Debye layer thickness $\lambda_{*}$ is given by 
\begin{equation}
\lambda_{*}^{2}=\frac{\varepsilon_{*}k_{B}T}{2c_{*}e^{2}} .
\label{lambdadef}
\end{equation}


Dimensionless groups that appear are 
\begin{subeqnarray}
\tau=\frac{k_{d*}a}{U_{c}} \gg 1, \ \ \ \alpha= \frac{c_{*}}{s_{*}} \gg 1, \ \ \ \epsilon=\frac{\lambda_{*}}{a} \ll 1, \hspace{2.5mm} \\
\Delta=\frac{\sigma}{\mu_{*}U_{c}}, \ \ \  \nu=\frac{\varepsilon_{*}(k_{B}T/e)^{2}}{\mu_{*}D_{*}} , \ \  \mbox{and} \ \ \Psi = \frac{e\Phi_{\infty}}{k_{B}T} . 
\label{paramdef}
\end{subeqnarray}
The role of $\tau$ and $\alpha$ is considered below, in \S \ref{sec:SEL}.  Except for this, the analysis is carried out in the limit of small 
Debye layer thickness, $\epsilon\rightarrow 0$, with $\Delta, \nu$, and $\Psi$ all of order $O(1)$.  The quantity $\Delta$ is the ratio of the capillary 
velocity, $\sigma/\mu_{*}$ to the ion-diffusive velocity $U_{c}$.  The group $\nu\epsilon$ is the ratio of electrostatic stress to viscous stress, and is 
referred to either as the inverse of a Mason number or as an electrical Hartmann number.  If the Helmholtz-Smoluchowski slip velocity $U_{HS}$ is 
based on the thermal voltage, then $U_{HS}=\varepsilon_{*}(k_{B}T/e)^{2}/\mu_{*}a$, and the group $\nu\epsilon= U_{HS}/U_{c}$ is also the ratio of the 
two velocity scales.  The parameters $\epsilon, \Delta$, and $\nu$ are all fixed for a given choice of electrolytes and initial drop size.  In contrast, the 
parameter $\Psi$ is the ratio of the applied potential difference $\Phi_{\infty}$ to the thermal voltage, and is a control parameter.  

In the scaling regime of this study, the Peclet number $Pe=U_{c}a/D_{*}$ does not appear independently, since the velocity scale 
for $U_{c}$ implies that $Pe=\epsilon^{-1}$.  

The same variable and parameter names are now used to express the governing equations and boundary conditions in nondimensional form.  
The molecular ion flux $\boldsymbol{j}^{(\pm)}=c^{(\pm)}(\boldsymbol{v}^{(\pm)} - \boldsymbol{u})$ of (\ref{jdef0}), recast via (\ref{NPdimb}) and 
the relation $D^{(\pm)}=b^{(\pm)}k_{B}T$, is  
\begin{equation}
\boldsymbol{j}^{(\pm)} = - \epsilon D^{(\pm)}( \nabla c^{(\pm)} \pm c^{(\pm)}\nabla \phi ) \, , 
\label{jdef}
\end{equation} 
in nondimensional form.  Conservation of ions (\ref{NPdim}) and salt (\ref{sdim}), together with the expression (\ref{R}) for the ion kinetic rate 
term $\cal{R}$ and the incompressibility condition of (\ref{Stokesdim}) give 
\begin{eqnarray}
(\partial_{t}+\boldsymbol{u}\cdot\nabla)c^{(\pm)}-\epsilon D^{(\pm)}\nabla\cdot ( \nabla c^{(\pm)}\pm c^{(\pm)}\nabla \phi) = 
\tau \alpha^{-1} (k_{d}s-k_{a}c^{(+)}c^{(-)}) \, ,
\label{NPrate} \\[6pt]
(\partial_{t}+\boldsymbol{u}\cdot\nabla)s-\epsilon D_{s}\nabla^{2} s =  - \tau (k_{d}s-k_{a}c^{(+)}c^{(-)}) \, . \hspace{20mm}
\label{s}
\end{eqnarray}

The Poisson equation (\ref{Pdim}) is 
\begin{equation}
\epsilon^{2}\nabla^{2}\phi=-\frac{q}{\varepsilon} \, , \ \ \mbox{where} \ \ q = \frac{c^{(+)}-c^{(-)}}{2} ,
\label{Peqn}
\end{equation}
so that $q$ is a normalized charge density.  The continuity and Stokes equation are 
\begin{equation}
\nabla \cdot \boldsymbol{u}=0,\ \ \mbox{and} \ \ -\nabla p + \mu \nabla^{2}\boldsymbol{u} + 
\varepsilon \nu \epsilon \nabla^{2}\phi\nabla\phi = 0 , 
\label{Stokes}
\end{equation}
where the body force is written in terms of the potential via (\ref{Efield}).

{\it \underline{Drop Boundary Conditions.}}  In nondimensional form, from (\ref{Efield}) and (\ref{jdef}), the interfacial boundary conditions (\ref{BC1dim}) 
on $S$ become 
\begin{eqnarray}
[\phi]=0, \ \ [\varepsilon \nabla \phi\cdot\boldsymbol{n}] = 0 \, , \ \mbox{and} \hspace{22mm}
\label{BC1a} \\ 
\boldsymbol{j}^{(\pm)} \cdot\boldsymbol{n}=0 \, , \ \mbox{or equivalently} \ ( \nabla c^{(\pm)}\pm c^{(\pm)}\nabla \phi)\cdot \boldsymbol{n}=0  \, .
\label{BC1b}
\end{eqnarray}
Continuity of velocity and the kinematic condition are formally unchanged, namely  
\begin{equation}
[\boldsymbol{u}]=0 \ \ \mbox{and} \ \ (\boldsymbol{u} - \frac{\partial\boldsymbol{x}_{s}}{\partial t})\cdot\boldsymbol{n}=0 \ \ \mbox{on} \ S . 
\label{BC2}
\end{equation}
The stress-balance boundary condition becomes 
\begin{eqnarray}
[(\boldsymbol{T}_{H} + \nu\epsilon\boldsymbol{T}_{M})\cdot\boldsymbol{n}] = \Delta(\kappa_{1}+\kappa_{2})\boldsymbol{n} 
            \ \ \mbox{on} \ S, \hspace{20mm} 
\label{sbal} \\
\mbox{where} \ \ \boldsymbol{T}_{H}= -p\boldsymbol{I}+2 \mu \boldsymbol{e} \ \ \mbox{and} \ \ 
\boldsymbol{T}_{M}= \varepsilon (\nabla \phi \otimes \nabla \phi-\frac{1}{2}|\nabla\phi|^{2}\boldsymbol{I}) \, .
\label{tensnondim}
\end{eqnarray}
The boundary conditions (\ref{G1}) for a Galvani potential (when $l^{(+)}\neq l^{(-)}$) are formally unchanged, namely 
\begin{equation}
c^{(+)}|_{\partial\Omega_{1}} = l^{(+)} c^{(+)}|_{\partial\Omega_{2}}  , \ \mbox{and} \  c^{(-)}|_{\partial\Omega_{1}} = l^{(-)} c^{(-)}|_{\partial\Omega_{2}}
  \ \ \mbox{on} \ S \, .
\label{G2}
\end{equation}

{\it \underline{Vesicle Boundary Conditions.}}  In nondimensional form the trans-membrane ion flux $\boldsymbol{j}^{(\pm)} \cdot\boldsymbol{n}$ is given by 
\begin{equation}
c^{(\pm)} (\boldsymbol{v}^{(\pm)}-\boldsymbol{u}) \cdot \boldsymbol{n} = - \epsilon g^{(\pm)} [ \ln c^{(\pm)} \pm  \phi ] \ \ \mbox{on} \ S \, ,
\label{ionfluxmem} 
\end{equation}
where the gating constant $g^{(\pm)}$ of (\ref{ionfluxdim2}) has been made nondimensional and scaled by $g_{*}=\frac{c_{*}D_{*}}{k_{B}Ta}$.  
The flow of solvent across the membrane is given by 
\begin{equation}
\boldsymbol{u} - \left.\frac{\partial\boldsymbol{X}}{\partial t}\right|_{\boldsymbol{\theta}} = 
                  - \, \epsilon \pi_{m} \, [ \frac{2\epsilon}{\nu} p - (c^{(+)}+c^{(-)}) ] \, \boldsymbol{n} \ \ \mbox{on} \ S \, ,
\label{solfluxmem}
\end{equation}
where the effective porosity $\pi_{m}$ of (\ref{solfluxdim}) has been made nondimensional and scaled by $\pi_{*}=\frac{D_{*}}{k_{B}Tc_{*}a}$.   These relations 
contrast with the boundary conditions (\ref{BC1b}) and (\ref{BC2}) for an ideally polarizable drop with a sharp impervious interface; the cell has a small 
$O(\epsilon)$ flux of ions and solvent across the membrane when it is semi-permeable to these species.  

The relations (\ref{capdim}) that express continuity of electric displacement across the membrane become 
\begin{equation}
\varepsilon \epsilon \nabla \phi\cdot\boldsymbol{n}|_{\partial\Omega_{1}} = \varepsilon \epsilon \nabla \phi\cdot\boldsymbol{n}|_{\partial \Omega_{2}} 
   = C_{m} [\phi]  \, . 
\label{capbc}
\end{equation}
Since a lipid bilayer or similar biomembrane has a thickness of the order of $5$nm to $8$nm, which is the same order of magnitude as the thickness 
of the neighboring Debye layers, the membrane thickness has been made nondimensional by $\lambda_{*}$ and the membrane capacitance per unit area, 
$C_{m}$ or $C_{m}^{0}$ of (\ref{capexd}), has been made nondimensional by $\varepsilon_{*}/\lambda_{*}$.  This accounts for the $\epsilon$-scaling in 
(\ref{capbc}).  

The stress-balance boundary condition is now 
\begin{equation}
[(\boldsymbol{T}_{H} + \nu \epsilon \boldsymbol{T}_{M})\cdot\boldsymbol{n}] = 2 \sigma H \boldsymbol{n} - \nabla_{s} \sigma  
      -  \kappa_{b} \left( 2H(H^{2}-K)  +  \nabla_{s}^{2} H \right) \boldsymbol{n} \, ,
\label{cellsbal} 
\end{equation}
where the tension in the membrane and the bending modulus have been made nondimensional by $\mu_{*}U_{c}$ and $\mu_{*}U_{c}a^{2}$ respectively, 
and the spontaneous curvature $C_{0}$ has been set to zero.  The membrane inextensibility constraint is formally unchanged, namely 
\begin{equation}
\nabla_{s}\cdot\left( \frac{\partial}{\partial t}\boldsymbol{X}_{s}(\boldsymbol{\theta}, t) \right) =0 \, . 
\label{inext}
\end{equation}

{\it \underline{Initial and Far-Field Conditions.}}  The initial conditions are 
\begin{equation} 
\phi(\boldsymbol{x}, 0^{+}) = - \Psi z, \ \boldsymbol{u}(\boldsymbol{x}, 0) = 0, \ \mbox{and} \ c^{(\pm)}(\boldsymbol{x}, 0) = c_{i} \ \mbox{on} \ \Omega_{i}, \ i=1,2\, , 
\label{ics}
\end{equation}
except that the first and third relations are revised for a drop when it has a nonzero Galvani potential, as considered in \S \ref{sec:Galvani}, and we have assumed a 
uniform applied field.  Here $c_{i}$ is the initial 
dimensionless ion concentration on $\Omega_{i}$ for $i=1,2$.  A drop is initially spherical with radius 1, 
whereas a vesicle has a known initial equilibrium configuration $S_{ref}$.  Far from a drop or vesicle, 
\begin{equation}
\phi(\boldsymbol{x}, t) \sim - \Psi z, \  \boldsymbol{u}(\boldsymbol{x}, t) \rightarrow 0, \ \mbox{and} \  
c^{(\pm)}(\boldsymbol{x}, t)\rightarrow \  c_{2} \ \mbox{as} \ |\boldsymbol{x}|\rightarrow \infty \ \mbox{for} \ t > 0\, . 
\label{far-field}
\end{equation}
The applied potential and constant electric field are related by $\Phi_{\infty}=E_{\infty}a$.  For a more general applied field $\phi$ approaches a specified solution 
$\phi_{\infty}$ of Laplace's equation that has magnitude $\Psi$.

\section{The strong electrolyte limit} 
\label{sec:SEL}

The two dimensionless groups $\tau$ and $\alpha$ that multiply the rate term $\cal{R}$ on the right hand side of equations (\ref{NPrate}) and (\ref{s}) are defined 
at (\ref{paramdef}a).  Of these, $\tau=k_{d*}a/U_{c}$ is the ratio of the dissociation rate of salt to the rate of Debye layer charge-up, which is 
taken to be large.  The group $\alpha=c_{*}/s_{*}$ is the ratio of the ambient concentration of the dissociated ions to that of the combined salt, which is large for a 
strong electrolyte.  

In the limit $\tau \rightarrow\infty$, equation (\ref{s}) implies that to leading order
\begin{equation}
k_{d}s-k_{a} c^{(+)}c^{(-)}=0 \, ,
\label{rateeqm1}
\end{equation}
that is, for all $\boldsymbol{x}$ and $t$ the ion and salt concentrations are at leading order equilibrium throughout both phases.   The right hand side of the 
ion conservation equation (\ref{NPrate}) is now of order $\tau/\alpha$ times a residual rate term that is of order $o(1)$.  Provided $\tau/\alpha$ is bounded, that is, 
provided $\alpha\rightarrow\infty$ as fast as or faster than $\tau\rightarrow\infty$ the right hand side of (\ref{NPrate}) is $o(1)$, which is 
sufficiently small for it to be neglected as a higher order effect in the small-$\epsilon$ analysis that is developed below.  

The conservation equations (\ref{NPrate}) and (\ref{s}) can now be simplified:  the $o(1)$ right hand side rate term of equation (\ref{NPrate}) 
for the ion concentrations $c^{(+)}$ and $c^{(-)}$ is omitted, and after the ion concentrations have been found the leading order salt concentration $s$ is given 
by (\ref{rateeqm1}), if required.  Equation (\ref{s}) for conservation of $s$ can then be omitted, since it contains no further information at the order of 
calculation of the model.  Instead of equations (\ref{NPrate}) and (\ref{s}), we have 
\begin{equation}
(\partial_{t}+\boldsymbol{u}\cdot\nabla)c^{(\pm)}-\epsilon D^{(\pm)}\nabla\cdot ( \nabla c^{(\pm)}\pm c^{(\pm)}\nabla \phi) = 0 \, .
\label{NP1}
\end{equation}

An equation for conservation of charge $q$ is given by forming the difference of equations (\ref{NP1}) for the ion species, or by forming the difference 
of equations 
(\ref{NPrate}), to give 
\begin{eqnarray}
(\partial_{t}+\boldsymbol{u}\cdot\nabla) q = \hspace{90mm} \nonumber \\
\frac{\epsilon}{2} \left( D^{(+)}\nabla^{2}c^{(+)} -  D^{(-)}\nabla^{2}c^{(-)} + D^{(+)}\nabla\cdot (c^{(+)}\nabla \phi) + D^{(-)}\nabla\cdot (c^{(-)}\nabla \phi) \right) 
\label{qcons}
\end{eqnarray}
which is exact.  The terms grouped on the right hand side represent charge diffusion.

\section{The outer regions away from the Debye layers}
\label{sec:outer}

In the outer regions, away from the Debye layers, the dependent variables are expanded in integer powers of $\epsilon$, so 
that for the ion concentrations and potential, respectively 
\begin{subeqnarray}
c^{(\pm)}&=&c_{0}^{(\pm)}+\epsilon c_{1}^{(\pm)}+\ldots \, , \slabel{outansatza}\\
\phi&=&\phi_{0}+\epsilon \phi_{1}+\ldots \, , \slabel{outansatzb}
\label{outansatz}
\end{subeqnarray}
with analogous notation for expansion of $\boldsymbol{u}$, $p$, and $q$.  

To a high degree of approximation the outer regions are charge neutral and the electrostatic body force in the momentum equation is zero.  
To see this, note that the Poisson equation (\ref{Peqn}) implies that in an outer approximation $q_{0} = q_{1}=0$, so that 
\begin{equation}
c_{0}^{(+)}=c_{0}^{(-)} \ \ \mbox{and} \ \  c_{1}^{(+)}=c_{1}^{(-)} 
\label{out1}
\end{equation}
for all $\boldsymbol{x}$ and $t$.  Then, at leading order the ion conservation equations (\ref{NP1}) imply that 
\begin{equation}
(\partial_{t}+\boldsymbol{u}_{0}\cdot\nabla)c_{0}^{(\pm)}=0 \, , 
\label{trans1}
\end{equation}
so that throughout the outer regions, from the first of relations (\ref{out1}), 
\begin{subeqnarray}
(\partial_{t}+\boldsymbol{u}_{0}\cdot\nabla)c_{0} = 0 \, , \hspace*{33mm} \slabel{out2a} \\
\mbox{where} \ \ c_{0}^{(+)} = c_{0}^{(-)}  \stackrel{\mathrm{def}}{=} c_{0} \ \mbox{is conserved on particle paths} \ 
        \frac{d\boldsymbol{x}}{dt}=\boldsymbol{u}_{0} \, . \slabel{out2b}  
\label{out2}
\end{subeqnarray}
Since the two ion concentrations are equal the indication of their charge has been omitted.  

The outer regions of $\Omega_{1}$ and $\Omega_{2}$  are partitioned into subdomains.  An outflow subdomain consists of fluid that, at any time $t\ge 0$, originates 
from or exits a Debye layer.  For want of better terminology we refer to its complement, which is non-empty at least for sufficiently early times, as an inflow subdomain.  
Notice that if the flow in the outer regions continually recirculates, an inflow subdomain can vanish at some time.  However, at the outer edges of the Debye 
layers, there are always outflow regions where fluid exits the layer to enter the bulk and complementary inflow regions where fluid enters the layer from the bulk.  
The inflow and outflow subdomains are three-dimensional volumes, whereas the inflow and outflow regions are surfaces.  

\begin{figure}
\begin{center}
\includegraphics[scale=0.20]{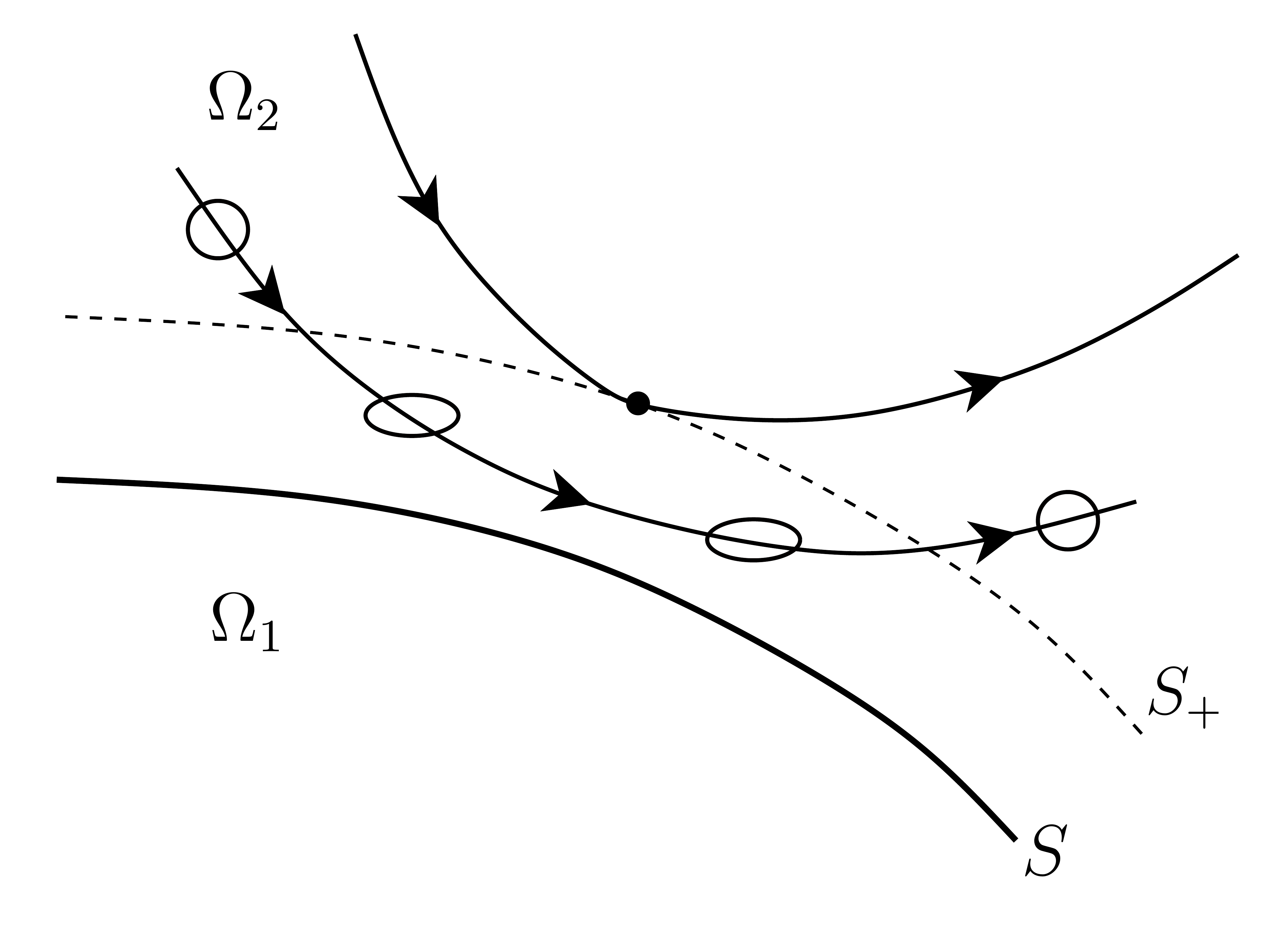} 
\end{center}
\caption{Sketch of characteristics or particle paths $\boldsymbol{x}$ such that $\frac{d\boldsymbol{x}}{dt}=\boldsymbol{u_{0}}$ showing inflow and outflow 
regions for the Debye layer in the exterior domain $\Omega_{2}$.  
The inflow and outflow regions are separated by the location of the marker ($\bullet$), which is such that $\partial_{n}v_{p}|_{S}=0$ as explained in \S \ref{sec:hydro0}.  
An impression is given of the deformation of a Lagrangian fluid volume as it enters and leaves the Debye layer.  
The counterpart in $\Omega_{1}$ is not shown.  
}  
\label{fig:sketchup}
\end{figure}

Figure \ref{fig:sketchup} gives an illustrative sketch of outflow and inflow regions neighboring the interface.  They are distinguished by the sign of the quantity 
$\partial_{n}v_{p}|_{S}$, which is introduced in  \S \ref{sec:hydro0} and is the rate of extension (for $\partial_{n}v_{p}|_{S}>0$) or contraction 
(for $\partial_{n}v_{p}|_{S}<0$) of a Lagrangian fluid line element that is normal to the interface.

On the inflow subdomains of $\Omega_{i}$, at sufficiently early times the far-field and initial conditions (\ref{far-field}) and (\ref{ics}) imply that (\ref{out2}) 
becomes 
\begin{equation}
c_{0} = c_{i} \ \mbox{is constant} \ \mbox{for} \ i=1,2 \, ,
\label{out3}
\end{equation}
where $c_{i}$ is the initial ambient ion concentration on $\Omega_{i}$.
Since $q_{1}=0$, equation (\ref{qcons}) for charge conservation at order $O(\epsilon)$ implies that $\nabla^{2}\phi_{0}=0$.  
From this, the Poisson equation (\ref{Peqn}) implies that $q_{2}=0$ also, and the result of charge neutrality at higher integer powers of $\epsilon$ begins 
to repeat or bootstrap:  since $\nabla^{2}\phi_{0}=0$ and $c_{0}^{(\pm)}$ is constant, ion conservation given by (\ref{NP1}) at order $O(\epsilon)$ 
implies that $(\partial_{t}+\boldsymbol{u}_{0}\cdot\nabla)c_{1}^{(\pm)}=0$.  The second of relations (\ref{out1}) with the far-field and initial 
conditions implies that $c_{1}^{(+)}$ and $c_{1}^{(-)}$ are equal and constant on each inflow subdomain.  This constant would usually be set to 
zero, but in any event, since $q_{2}=0$ equation (\ref{qcons}) for charge conservation at order $O(\epsilon^{2})$ now implies that $\nabla^{2}\phi_{1}=0$, 
et cetera.  

On the outflow subdomains of $\Omega_{i}$ relations (\ref{out1}) to (\ref{out2}) still hold, but the initial data for the transport equation (\ref{trans1}) is given 
by matching to the ion concentrations on outflow from the evolving Debye layers.  In contrast to (\ref{out3}), $c_{0}$ is then a function of $\boldsymbol{x}$ 
and $t$ in the Eulerian frame.  Similarly, if regions of re-entrant flow develop, when fluid re-enters a Debye layer it carries a value of 
$c_{0}(\boldsymbol{x}, t)$ equal to the ion concentration that it held at its previous exit.  Since $q_{1}=0$, equation (\ref{qcons}) for conservation 
of charge at order $O(\epsilon)$ now implies a constraint or consistency condition between the ion concentration $c_{0}$ and the potential $\phi_{0}$, which is 
\begin{equation}
(D^{(+)}-D^{(-)})\nabla^{2}c_{0} + (D^{(+)}+D^{(-)}) \nabla \cdot (c_{0} \nabla \phi_{0} ) = 0 \, .
\label{outflowconsist}
\end{equation}

Without formal proof, we assume that $q_{2}=0$ everywhere, so that from (\ref{Peqn}) the potential satisfies the Laplace equation 
\begin{equation}
\nabla^{2}\phi_{0}=0
\label{Laplace}
\end{equation}
throughout the outer domains.  It follows that the equations of Stokes flow are also unforced in the outer domains.  That is  
\begin{equation}
\nabla \cdot \boldsymbol{u}_{0}=0 \, ,\ \ \mbox{and} \ \ -\nabla p_{0} + \mu \nabla^{2}\boldsymbol{u}_{0} = 0 \, . 
\label{Stokes2}
\end{equation}

\section{The Debye layers: equilibrium solution and hydrodynamics}
\label{sec:layers1}

\subsection{The intrinsic coordinate system}
\label{sec:intrinsic}

To resolve the dynamics of the Debye layers we introduce a local surface-fitted or intrinsic orthogonal curvilinear coordinate system.  This has 
tangential coordinates $\xi_{1}$ and $\xi_{2}$ that are aligned with the principal directions on $S$ and normal coordinate $n$, with $n=0$ on 
$S$ and $n>0$ in $\Omega_{2}$.   The origins of the Eulerian and intrinsic coordinate systems are $O$ and $O^{\prime}$, respectively, and 
the position vector $\boldsymbol{x}$ of a point $P$ in space relative to $O$ is written in the two coordinate systems as 
\begin{equation}
\boldsymbol{x}=\boldsymbol{X}(\xi_{1}, \xi_{2}, t) + n \boldsymbol{n}(\xi_{1}, \xi_{2}, t).  
\label{pdef}
\end{equation}
Here $\boldsymbol{x}=\boldsymbol{X}(\xi_{1}, \xi_{2}, t)$ is the parametric equation of $S$ and $\boldsymbol{n}$ is its outward unit normal.  
Since $\xi_{1}$ and $\xi_{2}$ are principal directions on $S$ they define an orthogonal coordinate system on it with associated unit tangent 
vectors $\boldsymbol{e}_{i}=\frac{1}{a_{i}}\frac{\partial \boldsymbol{X}}{\partial \xi_{i}}$, where 
$a_{i}=|\frac{\partial \boldsymbol{X}}{\partial \xi_{i}}|$ for $i=1, 2$.   With the convention that 
$\boldsymbol{n}=\boldsymbol{e}_{1}\times\boldsymbol{e}_{2}$, Rodrigues' formula implies that the principal curvatures $\kappa_{i}$ satisfy 
$\frac{\partial \boldsymbol{n}}{\partial \xi_{i}}=\kappa_{i}\frac{\partial \boldsymbol{X}}{\partial \xi_{i}}$, and the change in $\boldsymbol{x}$ 
corresponding to increments in the intrinsic coordinates with time fixed is 
\begin{equation}
d\boldsymbol{x}=l_{1}d\xi_{1}\, \boldsymbol{e}_{1}+l_{2}d\xi_{2}\, \boldsymbol{e}_{2}+dn \, \boldsymbol{n}, \ \ \mbox{where} \ \ 
l_{i}=a_{i}(1+n\kappa_{i}) \ \ (i=1, 2).
\label{dx}
\end{equation}
Expressions for vector differential operators and the rate of strain tensor in a general orthogonal curvilinear coordinate system can be found in, 
for example, \cite{Batchelor}, Appendix 2.

Dependent variables within the Debye layers are denoted by upper case letters, with terms in an expansion in integer powers of $\epsilon$ 
denoted by subscripts, so that for the ion concentrations $c^{(\pm)}=C^{(\pm)}=C_{0}^{(\pm)}+\epsilon C_{1}^{(\pm)}+\ldots$ and for the potential 
$\phi=\Phi= \Phi_{0}+\epsilon \Phi_{1}+\ldots$, for example.  The analysis of the Debye layers is based on a local normal coordinate $N$ 
defined by 
\begin{equation}
n=\epsilon N  \, , 
\label{Ndef}
\end{equation}
where $N=O(1)$ as $\epsilon\rightarrow 0$.  

\subsection{The Gouy-Chapman solution}
\label{sec:pnp0}

The leading order ion concentrations and potential are given by the Gouy-Chapman solution.  Reviews of this are given by, for example, \cite{Russel_etal} 
and \cite{Hunter}, and it is summarized here because it is the foundation for the analysis that follows.  In the present context, all dependent variables depend 
parametrically on the tangential coordinates $\xi_{1}$ and $\xi_{2}$, and on time $t$.  

In terms of local variables, at leading order, the ion conservation equations (\ref{NP1}) and the interface condition (\ref{BC1b}) for a drop, 
or (\ref{ionfluxmem}) with (\ref{jdef}) for a vesicle, imply that
\begin{subeqnarray}
\partial_{N} (\partial_{N}C_{0}^{(\pm)}\pm C_{0}^{(\pm)}\partial_{N}\Phi_{0})=0 \ \ N \neq 0, 
\slabel{GC1a} \\
\mbox{with} \ \ \partial_{N}C_{0}^{(\pm)}\pm C_{0}^{(\pm)}\partial_{N}\Phi_{0} =0 \ \ \mbox{on} \ S . \hspace{2mm}
\slabel{GC1b}
\label{GC1}
\end{subeqnarray}
An integration immediately gives 
\begin{equation}
\partial_{N}C_{0}^{(\pm)}\pm C_{0}^{(\pm)}\partial_{N}\Phi_{0} =0 \ \ \mbox{for all} \ N \, .
\label{GC2}
\end{equation}
This has a simple interpretation in terms of the molecular ion flux (\ref{jdef}), which has local expansion 
$\boldsymbol{j}^{(\pm)}=\boldsymbol{J}_{0}^{(\pm)}+\epsilon\boldsymbol{J}_{1}^{(\pm)} + O(\epsilon^{2})$ in the Debye layers, and where 
the leading order flux $\boldsymbol{J}_{0}^{(\pm)}$ is in the normal direction.  Equation (\ref{GC2}) states that $\boldsymbol{J}_{0}^{(\pm)}$ is zero throughout 
the layers, with diffusion and electromigration of ions in equilibrium at this level of approximation.

The requirement of matching with the outer 
regions is that 
\begin{equation}
C_{0}^{(\pm)} \rightarrow 
\left\{
\begin{array}{cc}
c_{0}|_{S_{+}} & \mbox{as} \ N\rightarrow \infty \\
c_{0}|_{S_{-}} & \mbox{as} \ N\rightarrow -\infty
\end{array}
\right. 
\ \ \mbox{and} \ \ \Phi_{0}\rightarrow 
\left\{
\begin{array}{cc}
\phi_{0}|_{S_{+}} & \mbox{as} \ N\rightarrow \infty \\
\phi_{0}|_{S_{-}} & \mbox{as} \ N\rightarrow -\infty
\end{array}
\right. ,
\label{match1}
\end{equation}
where an $S_{+}$ or $S_{-}$ subscript denotes, respectively, the limit of a quantity in the outer regions as $S$ is approached from the exterior of the drop  
(as $n\rightarrow 0^{+}$) or from the interior (as $n\rightarrow 0^{-}$).  We have just shown at equation (\ref{out3}) that the ion concentration 
$c_{0}|_{S_{\pm}}=c_{i}$ ($i=1,2$) is constant over regions where the flow enters the Debye layers, at least at early times, but it is to be determined and 
satisfies (\ref{out2}) elsewhere.  

Integration of (\ref{GC2}) with the matching conditions (\ref{match1}) gives the Boltzmann 
distribution between the ion concentrations and the potential, namely
\begin{subeqnarray}
C_{0}^{(+)}=c_{0}|_{S_{\pm}}e^{-\eta_{0}} \, , \ \ \  C_{0}^{(-)}=c_{0}|_{S_{\pm}}e^{\eta_{0}} \, , 
\slabel{C0a} \\
\mbox{where} \ \ \eta_{0}= 
\left\{
\begin{array}{cc}
\Phi_{0}-\phi_{0}|_{S_{+}} & \mbox{for} \ N>0 \\
\Phi_{0}-\phi_{0}|_{S_{-}} & \mbox{for} \ N<0 
\end{array}
\right. .
\slabel{epb}
\label{C0ep}
\end{subeqnarray}
Here $\eta_{0}$ is the first term in the expansion of $\eta=\Phi-\phi|_{S_{\pm}}$, which is the excess potential in the Debye layers.  

From equation (\ref{Peqn}), the charge density $q$ in the layers is therefore given at leading order by $Q_{0}=-c_{0}|_{S_{\pm}}\sinh \eta_{0}$,
and the potential $\Phi_{0}$ satisfies the Poisson-Boltzmann equation 
\begin{equation}
\varepsilon\partial_{N}^{2}\Phi_{0}= c_{0}|_{S_{\pm}} \sinh  (\Phi_{0}-\phi_{0}|_{S_{\pm}}).  
\label{PB}
\end{equation}
This can be integrated once on multiplying by $\partial_{N}\Phi_{0}$ and using the matching conditions (\ref{match1}) to find 
\begin{equation}
\varepsilon (\partial_{N}\Phi_{0})^{2}=2 c_{0}|_{S_{\pm}} (\cosh (\Phi_{0}-\phi_{0}|_{S_{\pm}})-1) .
\label{1stint}
\end{equation}
For a drop, the first of the interfacial boundary conditions (\ref{BC1a}), that $[\phi]=0$, states that the potential is continuous across $S$ with a surface value 
denoted by $\lim_{N\rightarrow 0^{\pm}}\Phi_{0}=\Phi_{0}|_{S}$.  A second integration then gives the Gouy-Chapman solution in the form  
\begin{equation}
\tanh \frac{1}{4}(\Phi_{0}-\phi_{0}|_{S_{\pm}}) = \tanh \frac{1}{4} (\Phi_{0}|_{S}-\phi_{0}|_{S_{\pm}}) e^{{\mp}\sqrt{\left( c_{0}|_{S_{\pm}}/\varepsilon \right)} N}, 
\label{GC}
\end{equation}
where the upper choice of signs hold for $N>0$ and the lower choice for $N<0$.  For a vesicle, its membrane capacitance implies that there can be a non-zero 
jump in the potential across the vesicle membrane, with 
$\lim_{N\rightarrow 0^{-}}\Phi_{0} \equiv \Phi_{0}|_{\partial\Omega_{1}} \neq \lim_{N\rightarrow 0^{+}}\Phi_{0} \equiv \Phi_{0}|_{\partial\Omega_{2}}$.  This gives 
the Gouy-Chapman solution 
\begin{subeqnarray}
\tanh \frac{1}{4}(\Phi_{0}-\phi_{0}|_{S_{+}}) = & \tanh \frac{1}{4} (\Phi_{0}|_{\partial\Omega_{2}}-\phi_{0}|_{S_{+}}) \, e^{- \sqrt{\left( c_{0}|_{S_{+}}/\varepsilon \right)} N}  
\ \ \mbox{for} \ N>0 \, ,  \slabel{GCva} \\
\tanh \frac{1}{4}(\Phi_{0}-\phi_{0}|_{S_{-}}) = & \! \! \tanh \frac{1}{4} (\Phi_{0}|_{\partial\Omega_{1}}-\phi_{0}|_{S_{-}}) \, e^{ \sqrt{\left( c_{0}|_{S_{-}}/\varepsilon \right)} N} 
\ \ \ \mbox{for} \ N <0 \, . 
  \slabel{GCvb}
  \label{GCv}
\end{subeqnarray}

With the same convention for the choice of signs as in (\ref{GC}), the branch of the square root of equation (\ref{1stint}) is found by considering 
the behavior as $N\rightarrow \pm \infty$, which gives 
\begin{equation}
\partial_{N}\Phi_{0}=\mp 2 \sqrt{\frac{c_{0}|_{S_{\pm}}}{\varepsilon}} \sinh \frac{(\Phi_{0}-\phi_{0}|_{S_{\pm}})}{2} . 
\label{dNPhi0}
\end{equation}  
At this point we introduce the $\zeta$-potential for each layer.  This is the excess potential $\eta$, which is defined after equations (\ref{C0ep}), evaluated as 
$S$ is approached from the exterior or interior phase.  For a drop, with its continuous surface potential $\Phi_{0}|_{S}$ at $S$, the layer 
$\zeta$-potentials  are therefore 
\begin{eqnarray}
\zeta_{0+} &= & \Phi_{0}|_{S}-\phi_{0}|_{S_{+}} \, ,\nonumber \\
\zeta_{0-}  &= & \Phi_{0}|_{S}-\phi_{0}|_{S_{-}} \, . 
\label{zetapsD}
\end{eqnarray}
The second of the drop interfacial boundary conditions (\ref{BC1a}), which is the statement of Gauss's law across the interface, then implies the relation 
\begin{equation}
\sqrt{\varepsilon_{2} c_{0}|_{S_{+}}}\sinh \frac{\zeta_{0+}}{2} = - \sqrt{\varepsilon_{1} c_{0}|_{S_{-}}} \sinh \frac{\zeta_{0-}}{2} . 
\label{Gaussdrop}
\end{equation}
Recall that the electric permittivity, as a material parameter, is piecewise constant and may take different values $\varepsilon_{i}$ in the two 
distinct phases with $i=1, 2$.  For a vesicle, with its jump in potential across the membrane surface, the Debye layer $\zeta$-potentials are defined by 
\begin{eqnarray}
\zeta_{0+} &= & \Phi_{0}|_{\partial\Omega_{2}}-\phi_{0}|_{S_{+}} \nonumber \\
\zeta_{0-}  &= & \Phi_{0}|_{\partial\Omega_{1}}-\phi_{0}|_{S_{-}} , 
\label{zetapsV}
\end{eqnarray}
and the relation analogous to (\ref{Gaussdrop}) is given by the interfacial condition (\ref{capbc}), so that 
\begin{equation}
\sqrt{\varepsilon_{2} c_{0}|_{S_{+}}}\sinh \frac{\zeta_{0+}}{2} = - \sqrt{\varepsilon_{1} c_{0}|_{S_{-}}} \sinh \frac{\zeta_{0-}}{2} = 
    - \frac{C_{m}}{2} (\Phi_{0}|_{\partial\Omega_{2}}-\Phi_{0}|_{\partial\Omega_{1}}) \, . 
\label{Gausscell}
\end{equation}

A relation that will simplify an expression in the hydrodynamics of the Debye layers is given by taking the tangential derivative $\partial_{\xi_{i}}$ of 
equation (\ref{1stint}) and using (\ref{PB}), namely 
\begin{equation}
\partial_{N}^{2}\Phi_{0}\partial_{\xi_{i}}\Phi_{0} - \partial_{N}\Phi_{0}\partial^{2}_{\xi_{i}N}\Phi_{0} = 
          \partial_{N}^{2}\Phi_{0}\partial_{\xi_{i}}\phi_{0}|_{S_{\pm}} - \frac{(\partial_{N}\Phi_{0})^{2}}{2} \partial_{\xi_{i}} \ln c_{0}|_{S_{\pm}}      .
\label{id}
\end{equation}

The charge per unit area contained within each Debye layer can be found by integrating the Poisson equation (\ref{Peqn}) with respect to $N$.  
Integration across each layer implies that at leading order the normalized charge density per unit area (i.e., with the $1/2$ normalization factor in the volume 
charge density $q$ included) is $\varepsilon_{2}\partial_{N}\Phi_{0}|_{\partial\Omega_{2}}$ for $N>0$ (i.e., in $\Omega_{2}$) and is 
$- \varepsilon_{1}\partial_{N}\Phi_{0}|_{\partial\Omega_{1}}$ for $N<0$ (i.e., in $\Omega_{1}$). 
It follows from relations (\ref{BC1a}) for a drop and (\ref{capbc}) for a vesicle that the layer charges are equal and opposite for all time $t>0$ under dynamic 
conditions, that is, after the applied electric field is imposed, as well as under the $t<0$ equilibrium conditions of a Galvani potential.  

From equation (\ref{dNPhi0}), the amount of the leading order normalized charge density in the exterior layer, for example, is given by minus two times: the 
term on the left hand side of equation (\ref{Gaussdrop}) for a drop, and the term on the far left hand side of (\ref{Gausscell}) for a vesicle.  The charge densities 
are of opposite sign to the layer $\zeta$-potentials.

\subsection{Reduced expressions for the equilibrium Galvani and $\zeta$-potentials for a drop}
\label{sec:Galvani}

The equilibrium interface conditions (\ref{G2}) for a drop, together with the Boltzmann distribution (\ref{C0ep}) and the definition of the $\zeta$-potentials 
(\ref{zetapsD}), give the relations 
\begin{equation}
c_{0}|_{S_{-}} = l^{(+)} c_{0}|_{S_{+}} e^{[\phi_{0}]^{S_{+}}_{S_{-}}} =  l^{(-)} c_{0}|_{S_{+}} e^{-[\phi_{0}]^{S_{+}}_{S_{-}}} 
\label{gp1}
\end{equation}
between the ambient ion concentrations immediately outside the Debye layers $c_{0}|_{S_{\pm}}$, the jump in potential across 
the outer edges of the Debye 
layer pair $[\phi_{0}]^{S_{+}}_{S_{-}}$, and the ion partition coefficients $l^{(\pm)}$.  A re-arrangement implies that 
\begin{equation}
[\phi_{0}]^{S_{+}}_{S_{-}} =\ \frac{1}{2} \ln \left(\frac{l^{(-)}}{l^{(+)}}\right) \, ,
\label{gp2}
\end{equation}
where the equilibrium potential jump $[\phi_{0}]^{S_{+}}_{S_{-}}$ is the Galvani potential, and the ion concentrations are related by 
\begin{equation}
c_{0}|_{S_{-}}=c_{0}|_{S_{+}} \sqrt{l^{(+)}l^{(-)}} \, .
\label{gp3}
\end{equation} 

Expressions for the separate layer $\zeta$-potentials can be found when the relations (\ref{gp1}) are rewritten in terms of $\zeta_{0+}$ and $\zeta_{0-}$, 
and the pair $c_{0}|_{S_{-}}, \zeta_{0-}$ is eliminated in favor of the pair $c_{0}|_{S_{+}}, \zeta_{0+}$ (or vice-versa) in equation (\ref{Gaussdrop}).  
This gives 
\begin{equation}
\zeta_{0+} = \ln \left( \frac{1+ \sqrt{\frac{\varepsilon_{1} l^{(+)}}{\varepsilon_{2}}}}{1+ \sqrt{\frac{\varepsilon_{1} l^{(-)}}{\varepsilon_{2}}}} \right) \ \ \mbox{and} \ \ 
\zeta_{0-} = \ln \left( \frac{1+ \sqrt{\frac{\varepsilon_{2}}{\varepsilon_{1}l^{(+)}}}}{1+ \sqrt{\frac{\varepsilon_{2}}{\varepsilon_{1}l^{(-)}}}} \right) \, .
\label{Galvanizetas}
\end{equation}
It follows that when $l^{(-)}>l^{(+)}$ the exterior layer has $\zeta$-potential $\zeta_{0+}<0$ and carries net positive charge, while the interior layer has 
$\zeta_{0-}>0$ and net negative charge.  The converse holds when $l^{(-)}<l^{(+)}$.

\subsection{Debye Layer hydrodynamics}
\label{sec:hydro0}

The hydrodynamics in the Debye layers is described by considering the Eulerian fluid velocity $\boldsymbol{u}$ in the intrinsic frame of \S\ref{sec:intrinsic}.   
If $P(\xi_{1}, \xi_{2}, n)$ is fixed in space and $Q(\xi_{1}, \xi_{2}, 0)$ is the projection of $P$ onto $S$ 
in the normal direction, then the fluid velocity $\boldsymbol{u}$ at $P$ is written in terms of its tangential and normal projections in the intrinsic frame as 
\begin{equation}
\boldsymbol{u}= \boldsymbol{u}_{t} + u_{p}\boldsymbol{n} \, , \ \ \mbox{where} \ \boldsymbol{u}_{t}=u_{t1}\boldsymbol{e}_{1}+u_{t2}\boldsymbol{e}_{2} \, . 
\label{ua}
\end{equation}
In the Debye layers it is useful to compare the fluid velocity $\boldsymbol{u}$ at $P$ to the fluid velocity on the interface at $Q$, which in terms of its tangential 
and normal projections is $\boldsymbol{u}_{s}+u_{n}\boldsymbol{n}$.  The difference between the fluid velocity at $P$ and $Q$ is denoted similarly by 
$\boldsymbol{v}_{t} + v_{p} \boldsymbol{n}$, so that 
\begin{subeqnarray}
u_{p}&=&u_{n}+v_{p} , \slabel{uta} \\
\label{up}
u_{ti}&=&u_{si}+v_{ti} , \ \ i=1,2. \slabel{utb}
\label{ut}
\end{subeqnarray}

The interfacial velocity components $u_{n}$ and $u_{si}$ are necessarily independent of the normal coordinate $n$, and the dependence of the remaining components 
on $n$ expresses the shear within the Debye layers.  We find below that the relative velocity components $v_{ti}$ and $v_{p}$ are all of order $O(\epsilon)$ there. 

The scale of the normal component $v_{p}$ follows from the incompressibility condition $\nabla \cdot \boldsymbol{u}=0$, so that temporarily we set 
$v_{p}=\epsilon W_{0}+ O(\epsilon^{2})$.  In the intrinsic frame, the exact expression of the incompressibility condition is 
\begin{equation}
\partial_{\xi_{1}}(l_{2}u_{t1})+\partial_{\xi_{2}}(l_{1}u_{t2})+\partial_{n}(l_{1}l_{2}u_{p})=0 . 
\label{divexact}
\end{equation}
In the layers the dependent variables and the scale factors $l_{i}$ of (\ref{dx}) depend on the local normal coordinate $N=n/\epsilon$.  So that, with the 
convention that terms in a local expansion of a dependent variable are written in upper case, 
i.e., $u_{ti}=U_{ti0}+\epsilon U_{ti1}+O(\epsilon^{2})$ etc., the leading order expression of incompressibility is 
\begin{equation}
\frac{1}{a_{1}a_{2}} \left( \partial_{\xi_{1}}(a_{2}U_{t10})+\partial_{\xi_{2}}(a_{1}U_{t20}) \right) + 
    (\kappa_{1}+\kappa_{2})u_{n0} + \partial_{N}W_{0}=0 .
\label{locomp}
\end{equation}

We now turn to local expansion of the forced Stokes equation from equations (\ref{Stokes}).  When written in terms of 
$N$, the electrostatic body force has a normal component that is $O(\epsilon^{-2})$, and this can only be balanced by a normal 
pressure gradient of the same order.  The local expansion for the pressure is therefore
\begin{equation}
p=\frac{P_{-1}}{\epsilon}+P_{0} + \ldots \, . 
\label{Pexp}
\end{equation}
With the differential operators expressed in the local intrinsic coordinate frame, an estimate for the normal and tangential components 
of each term in the Stokes equation can be found, and this is shown in Table \ref{tab:termest}.

\begin{table}
\rule{165mm}{0.2mm}
  \begin{center}
\def~{\hphantom{0}}
  \begin{tabular}{ccccc}
                                    &   $-\nabla p$  & $+\mu \nabla^{2}\boldsymbol{u}$  &  $+\varepsilon \nu \epsilon \nabla^{2}\phi\nabla\phi$  & $=0$ \\[3pt]
 tangential component & $O(\epsilon^{-1})$ &  $O(\epsilon^{-2})$                 &  $O(\epsilon^{-1})$ &  \\
 normal component     & $O(\epsilon^{-2})$ & $O(\epsilon^{-1})$ or  $O(1)$ & $O(\epsilon^{-2})$  &  \\
  \end{tabular}
  \caption{Order of magnitude estimate for the terms in the Stokes equation.  The estimate for the normal component of the viscous force is revised 
  after considering the tangential component, as explained in the text.}
  \label{tab:termest}
  \end{center}
\rule{165mm}{0.2mm}
\end{table}

The leading order tangential component of the Stokes equation, at order $O(\epsilon^{-2})$, gives 
\begin{equation}
\partial_{N}^{2}U_{ti0}=0 , \ \ \ i=1,2 \, .  
\label{St1}
\end{equation}
Since a homogeneous solution for the tangential velocity $U_{i0}$ that is linear in $N$ can not match with the $O(1)$ velocity field away from 
the interface, the leading order fluid velocity in the layer has a plug flow profile, and is equal to the tangential velocity at the interface, viz. 
 \begin{equation}
 U_{ti0}=u_{si0}(\xi_{1}, \xi_{2}, t) , \ \ \ i = 1,2 \, .  
 \label{St2}
 \end{equation}
 In other words, there is no mechanism to support a boundary layer profile in the tangential velocity.
  
As a consequence of (\ref{St2}) the normal component of the viscous force in the Stokes equation is smaller than first estimated, and is $O(1)$ as shown 
revised in Table \ref{tab:termest}.  Further, all terms of equation (\ref{locomp}) are now seen to be independent of the local normal coordinate $N$, so that 
integration to find $v_{p}=\epsilon W_{0}+O(\epsilon^{2})$ is straightforward.

To find $v_{p}$, note first that the $\xi$-derivative terms of (\ref{locomp}) are the surface divergence of the tangential fluid velocity on the interface 
$\boldsymbol{u}_{s}$, at leading order, i.e., $\nabla_{s}\cdot \boldsymbol{u}_{s0}$.  For a drop with an impervious interface the relative velocity $v_{p}$ 
is zero on the interface $S$, where $N=0$, so that integration of (\ref{locomp}) with respect to $N$ gives 
\begin{subeqnarray}
v_{p} (\xi_{1}, \xi_{2}, N, t) = \epsilon \, \partial_{n}v_{p}|_{S} N  + O(\epsilon^{2}) \, , \hspace{6mm}  \slabel{v_p1a} \\ 
\mbox{where} \ \ \partial_{n}v_{p}|_{S} = - (\nabla_{s} \cdot \boldsymbol{u}_{s0}+(\kappa_{1}+\kappa_{2})u_{n0}) \, . \slabel{v_p1b} 
\label{v_p1}
\end{subeqnarray}
For a semi-permeable vesicle membrane the relative velocity $v_{p}$ on the interface is small, of order $O(\epsilon)$, and given by (\ref{solfluxmem}).  
Integration of (\ref{locomp}) with respect to $N$ then gives 
\begin{subeqnarray}
v_{p} (\xi_{1}, \xi_{2}, N, t) = \epsilon (v_{p}|_{S} +  \partial_{n}v_{p}|_{S} N ) + O(\epsilon^{2}) \, , \slabel{v_p2a} \\
\mbox{where} \ \ v_{p}|_{S} = - \pi_{m} [\frac{2}{\nu} P_{-1} - (C_{0}^{(+)} + C_{0}^{(-)}) ] \, . \slabel{v_p2b} 
\label{v_p2}
\end{subeqnarray}
The notation in equations (\ref{v_p2a}) and (\ref{v_p2b}) differs slightly from that elsewhere in this study in the expression of an $\epsilon$-scaling: 
equation (\ref{v_p2a}) gives the leading order value of $v_{p}$ on the interface as $v_{p}(N=0) = \epsilon v_{p}|_{S}$, where $v_{p}|_{S}$ of (\ref{v_p2b}) is 
order $O(1)$.  This choice will ease notation later. The derivative term $\partial_{n}v_{p}|_{S}$ is also order $O(1)$. 

The expression (\ref{v_p1b}) for $\partial_{n}v_{p}|_{S}$ holds for flow quantities on and immediately adjacent to $S$ for both the drop 
and vesicle, and has a simple interpretation in terms of flow field kinematics for incompressible flow.  The group of terms 
$\nabla_{s} \cdot \boldsymbol{u}_{s0}+(\kappa_{1}+\kappa_{2})u_{n0}$ on the right hand side of (\ref{v_p1b}) is the surface divergence of 
all components of the interfacial fluid velocity, $\boldsymbol{u}_{s0}+ u_{n0}\boldsymbol{n}$, including the normal velocity of $S$; see for example \cite{Weatherburn} 
and note the different convention for the sign of the principal curvatures.  Given the local conservation of volume of a fluid element and the definition of $v_{p}$ at 
(\ref{uta}), $\partial_{n}v_{p}|_{S}$ is the rate of extension or contraction of an infinitessimal fluid line element normal to the interface.  
   
The normal component of the Stokes equation, to leading order, implies that 
\begin{equation}
-\partial_{N}P_{-1}+ \varepsilon\nu\,\partial_{N}^{2}\Phi_{0}\partial_{N}\Phi_{0}=0 ,
\label{plo}
\end{equation}
and the solution that tends to zero as $N\rightarrow \pm \infty$, which confines the large $O(\epsilon^{-1})$ pressure term to the Debye layers, is 
\begin{equation}
P_{-1}=\frac{\varepsilon\nu}{2}(\partial_{N}\Phi_{0})^{2} .
\label{P-1}
\end{equation} 
At the next order, the normal component of the Stokes equation is 
\[
 -\partial_{N}P_{0} + \varepsilon \nu \left( \partial_{N}^{2}\Phi_{0}\partial_{N}\Phi_{1} + \partial_{N}^{2}\Phi_{1}\partial_{N}\Phi_{0} 
 +(\kappa_{1}+\kappa_{2}) (\partial_{N}\Phi_{0})^{2} \right) =0 \, ,
 \]
 which has solution 
 \begin{equation}
 P_{0} = \varepsilon \nu \left( \partial_{N} \Phi_{0} \partial_{N}\Phi_{1} +(\kappa_{1}+\kappa_{2})\int_{\pm \infty}^{N}(\partial_{N^{\prime}}\Phi_{0})^{2}dN^{\prime}\right) 
  + p_{0}|_{S_{\pm}}.
 \label{P0}
 \end{equation}
 Here, the terms in parenthesis decay exponentially as $N \rightarrow \pm \infty$ because of the exponential decay of the excess potential implied by 
 the Gouy-Chapman solution (\ref{GC}) or (\ref{GCv}), and $p_{0}|_{S_{\pm}}=\lim_{n\rightarrow 0^{\pm}}p_{0}$ is the pressure in the outer regions immediately 
 outside the Debye layers.  
 

It is useful to resolve the velocity profile in the Debye layers further by considering the tangential component of the Stokes equation at order $O(\epsilon^{-1})$, 
and this result is needed to evaluate the tangential viscous stress on both sides of the interface.  
We have found that the components $(u_{t1}, u_{t2}, u_{p})$ of the Eulerian velocity (\ref{ua}) are independent of $N$ at order $O(1)$, so that the expression 
for the tangential viscous force at order $O(\epsilon^{-1})$ simplifies.  The component of the Stokes equation in the $\xi_{1}$-direction is therefore 
\begin{equation}
-\frac{1}{a_{1}}\partial_{\xi_{1}}P_{-1} + \mu \partial_{N}^{2}U_{t11} + \varepsilon\nu  \partial_{N}^{2}\Phi_{0} \frac{1}{a_{1}}\partial_{\xi_{1}}\Phi_{0}=0 ,
\label{St3}
\end{equation}
with an analogous expression for the component in the $\xi_{2}$-direction.  If $P_{-1}$ is eliminated in favour of $\Phi_{0}$ using (\ref{P-1}) and the relation 
(\ref{id}) is used, we find that
\[
\partial_{N}^{2}U_{t11}=-\frac{\varepsilon\nu}{\mu a_{1}} \left(  \partial_{N}^{2}\Phi_{0}\partial_{\xi_{1}}\phi_{0}|_{S_{\pm}} 
                                                                                - \frac{(\partial_{N}\Phi_{0})^{2}}{2} \, \partial_{\xi_{1}} \! \ln c_{0}|_{S_{\pm}}   \right) \, . 
\] 
This has the solution, in terms of the excess potential $\eta_{0}=\Phi_{0}-\phi_{0}|_{S_{\pm}}$ in each layer,  
\begin{eqnarray}
U_{t11} = - \frac{\varepsilon\nu}{\mu a_{1}} \left( \eta_{0} \, \partial_{\xi_{1}}\phi_{0}|_{S_{\pm}} - 
                 4 \ln ( \cosh \frac{\eta_{0}}{4}) \, \partial_{\xi_{1}} \! \ln c_{0}|_{S_{\pm}} \right)  \nonumber \\\
                         + N \partial_{n}u_{10}|_{S_{\pm}} + u_{t11}|_{S_{\pm}} \, . \hspace{5mm} 
\label{St4}
\end{eqnarray}
As $N\rightarrow \pm \infty$ this satisfies the matching condition that $U_{t11} \sim N \partial_{n}u_{10}|_{S_{\pm}} + u_{t11}|_{S_{\pm}}$, and the terms in 
parenthesis tend to zero exponentially.  The  tangential fluid velocity on the interface at order $O(\epsilon)$,  $u_{s11}$, will remain undetermined but 
an expression for the $O(\epsilon)$ jump in tangential velocity across the combined Debye layer pair can be found by taking the limit of (\ref{St4}) as 
$N\rightarrow 0^{\pm}$.  In constructing (\ref{St4}) the antiderivative $\varepsilon\int (\partial_{N}\Phi_{0})^{2} /2 \, dN$ as found after equation (\ref{10}) 
has been used, from which a second integration follows.  


Consideration of the stress-balance boundary condition, which is given by (\ref{sbal}) for a drop or (\ref{cellsbal}) for a vesicle, together with the expressions 
(\ref{tensnondim}) for the stress tensors, confirms that there is no net traction on either side of the interface at order $O(\epsilon^{-1})$.  The hydrodynamic and electrostatic tractions 
separately each have an order $O(\epsilon^{-1})$ component in the normal direction, but these cancel as a consequence of (\ref{P-1}).  This point is taken 
up in Appendix \ref{sec:appB}.

\subsection{Reduced expressions for the trans-membrane ion flux and solvent osmotic flow speed for a vesicle}
\label{sec:hydro0a}

The trans-membrane ion flux, $\boldsymbol{j}^{(\pm)}\cdot\boldsymbol{n}$ on $S$, is given in terms of the jump in the electrochemical potential across the 
membrane by equation (\ref{ionfluxmem}), which we write here as 
\begin{equation}
\boldsymbol{j}^{(\pm)} \cdot \boldsymbol{n} = - \epsilon g^{(\pm)} [ \ln c^{(\pm)} \pm  \phi ] \ \ \mbox{on} \ S \, .
\label{ionfluxmem2} 
\end{equation}
When the flux $\boldsymbol{j}^{(\pm)}$ within the Debye layers is given the local expansion 
$\boldsymbol{j}^{(\pm)}=\boldsymbol{J}_{0}^{(\pm)}+\epsilon\boldsymbol{J}_{1}^{(\pm)}+O(\epsilon^{2})$ 
we have the fundamental result that $\boldsymbol{J}_{0}$ is zero, as noted below equation (\ref{GC2}).   Then, when the electrochemical potential on the 
right hand side of (\ref{ionfluxmem2}) is written in terms of the local variables $C_{0}^{(\pm)}$ and $\Phi_{0}$, the Boltzmann distribution (\ref{C0ep}) 
implies that 
\begin{equation}
\boldsymbol{J}_{1}^{(\pm)}\cdot\boldsymbol{n} = -g^{(\pm)} \left( \ln (\frac{c_{0}|_{S_{+}}}{c_{0}|_{S_{-}}}) \pm (\phi_{0}|_{S_{+}}-\phi_{0}|_{S_{{-}}}) \right) 
      \ \ \mbox{on} \ S \, .
\label{ionfluxmem3}
\end{equation}
This is written in a more compact form as 
\begin{equation}
j|_{S}^{(\pm)} = - g^{(\pm)} [ \ln c_{0} \pm \phi_{0} ]^{S_{+}}_{S_{-}} \ \ \mbox{on} \ S \, , 
\label{jflux}
\end{equation}
where $j|_{S}^{(\pm)}$ denotes $\boldsymbol{J}_{1}^{(\pm)}\cdot\boldsymbol{n}$ evaluated on $S$, i.e., when $N=0$\, and $[\, \cdot \,]^{S_{+}}_{S_{-}}$ denotes 
the jump in a quantity across the outer edges of the Debye layer pair.

To interpret this we note that, in the Debye layers, the flux $\boldsymbol{J}_{0}$ being zero and the Boltzmann distribution together imply that the electrochemical 
potential is independent of $N$ to leading order, so that its jump across the membrane in (\ref{ionfluxmem2}) is equal to the jump across the outer edges of the 
layers in (\ref{jflux}).  

Similarly, for a semi-permeable membrane the scaled osmotic flow speed $v_{p}|_{S}$ of equations (\ref{v_p2}) can be written in terms of quantities immediately 
outside or at the outer edges of the Debye layers.  Inside the layers, the absolute pressure is given by (\ref{P-1}) with (\ref{1stint}) while the dissolved salt concentration 
is given by (\ref{C0ep}).  It turns out that their difference, which is the partial pressure of the solvent, is independent of $N$, and the 
expression for the jump across the membrane surface simplifies, to give 
\begin{equation}
v_{p}|_{S} = 2\pi_{m} [ c_{0} ]^{S_{+}}_{S_{-}} \, ,
\label{vps}
\end{equation}
which is proportional to the difference in the total ion concentrations at the outer edges of the layers and is independent of the layer $\zeta$-potentials.

\section{Ion transport in the Debye layers}
\label{sec:ion_transport}

The transport of ions in the Debye layers is determined by a local analysis of the ion conservation equations (\ref{NP1}) at order $O(1)$, for which the 
material derivative is needed.  The change of variables from the Eulerian to the intrinsic coordinate system gives the exact result that 
\begin{equation}
(\partial_{t}+\boldsymbol{u}\cdot\nabla) \mapsto \partial_{t}+\boldsymbol{v}_{t}\cdot\nabla_{t} + v_{p}\partial_{n} \, ,
\label{DDt_exact}
\end{equation}
which holds for all $n$.  On the right hand side, the time-derivative is taken in the moving frame, i.e. with the intrinsic coordinates 
$(\xi_{1}, \xi_{2}, n)$ fixed, and $v_{p}$ is the normal relative velocity introduced at equation (\ref{uta}).  The velocity $\boldsymbol{v}_{t}$ and tangential 
gradient $\nabla_{t}$ are given by 
\begin{equation}
\boldsymbol{v}_{t} = \boldsymbol{u}_{t} - \left. \frac{\partial \boldsymbol{X}}{\partial t} \right|_{\boldsymbol{\xi}} 
    - n \left. \frac{\partial \boldsymbol{n}}{\partial t} \right|_{\boldsymbol{\xi}} \ \ \mbox{and} \ \ 
    \nabla_{t} = \frac{\boldsymbol{e}_{1}}{l_{1}}\frac{\partial}{\partial \xi_{1}} + \frac{\boldsymbol{e}_{2}}{l_{2}}\frac{\partial}{\partial \xi_{2}} \, ,
\label{vt}
\end{equation}
where the tangential fluid velocity $\boldsymbol{u}_{t}$ was introduced at (\ref{ua}).  

The leading order expression for the material derivative in the Debye layers is given by setting $n=\epsilon N$ with $N=O(1)$ in (\ref{DDt_exact}) and 
(\ref{vt}) as $\epsilon \rightarrow 0$.  First, the time derivative in (\ref{DDt_exact}) is taken at fixed $(\xi_{1}, \xi_{2})$ with $n=0$.  In the expression for 
$\boldsymbol{v}_{t}$ given by (\ref{vt}) the tangential velocity $\boldsymbol{u}_{t}$ is approximated by its leading order value $\boldsymbol{u}_{s0}$ 
at the interface, as found at equation (\ref{St2}), and the term in $n$ is higher order.  Similarly, the tangential gradient $\nabla_{t}$ is approximated by the 
surface gradient $\nabla_{s}$, which is given by replacing $l_{i}$ by $a_{i}$ ($i=1, 2$).  The normal relative velocity $v_{p}$ in (\ref{DDt_exact}) is approximated 
by the first non-zero terms in its Taylor series, which are given in terms of the local coordinate $N$ at order $O(\epsilon)$ by equations (\ref{v_p1}) for a drop 
and by equations (\ref{v_p2}) for a vesicle.  

It follows that in the Debye layers the leading order expression for the material derivative is given by 
\begin{subeqnarray}
\partial_{t} +\boldsymbol{v}_{s0}\cdot\nabla_{s} + (v_{p}|_{S} + \partial_{n}v_{p}|_{S} N) \partial_{N} \, , \slabel{DDt_0a} \\
\mbox{where} \ \ \boldsymbol{v}_{s0} = \boldsymbol{u}_{s0} - \partial_{t}\boldsymbol{X}|_{\boldsymbol{\xi}} \, . \hspace{7mm} \slabel{DDt_0b} 
\label{DDt_0}
\end{subeqnarray}
Here, for $v_{p}$ we have used the slightly more general expression (\ref{v_p2a}) pertaining to a vesicle, which includes the expression (\ref{v_p1a}) for 
a drop when $v_{p}|_{S}$ is set to zero.  We note that $v_{p}|_{S}$, $\boldsymbol{v}_{s0}$ and $\partial_{n}v_{p}|_{S}$ are functions of $\xi_{1}, \xi_{2}$, 
and $t$ alone and are independent of $N$, and that $\boldsymbol{v}_{s0}$ and $\partial_{n}v_{p}|_{S}$ depend on interface data alone, see (\ref{v_p1b}).  

More details of the coordinate transformation and reduction have been given previously in the context of bulk-interface surfactant 
exchange at large bulk P\'{e}clet number, see  \cite{JCP_bubble, RPA2020}. 

From the ion conservation equations (\ref{NP1}), conservation at order $O(1)$ therefore requires that 
\begin{eqnarray}
D^{(\pm)} \partial_{N} \left( \partial_{N}C_{1}^{(\pm)} \pm (C_{0}^{(\pm)}\partial_{N}\Phi_{1} + C_{1}^{(\pm)} \partial_{N}\Phi_{0}) \right) 
                                                                                                                                                                              = \nonumber \hspace{25mm} \\
     \left( \partial_{t} +\boldsymbol{v}_{s0}\cdot\nabla_{s} + (v_{p}|_{S} + \partial_{n}v_{p}|_{S}N) \partial_{N} \right) C_{0}^{(\pm)}
\label{NP2}
\end{eqnarray}
for $N\neq 0$.  
As noted below equation (\ref{GC2}), the leading order molecular ion flux $\boldsymbol{J}_{0}^{(\pm)}$ is zero throughout the Debye layers, 
and as a result of this terms multiplied by the mean curvature are absent from the left hand side of (\ref{NP2}), while the terms in parenthesis 
there taken together with the diffusivity $D^{(\pm)}$ constitute the normal flux $-\boldsymbol{J}_{1}^{(\pm)}\cdot\boldsymbol{n}$ at the next order, as seen by 
expansion of (\ref{jdef}).

\subsection{Debye layer ion transport for a drop }
\label{sec:ITdrop}

For a drop with an interface that is impermeable to ions the boundary condition (\ref{BC1b}) implies that the normal flux term 
$\boldsymbol{J}_{1}^{(\pm)}\cdot\boldsymbol{n}$  vanishes at $N=0$, that is 
\begin{equation}
\partial_{N}C_{1}^{(\pm)} \pm (C_{0}^{(\pm)}\partial_{N}\Phi_{1} + C_{1}^{(\pm)} \partial_{N}\Phi_{0})=0 \ \ \mbox{on} \ S \, .  
\label{BC1hodrop}
\end{equation}
The interface is also impermeable to flow of the immiscible liquids that it separates, so that the relative normal fluid velocity $v_{p}|_{S}=0$, 
and an integration of (\ref{NP2}) gives 
\begin{eqnarray}
\partial_{N}C_{1}^{(\pm)} \pm (C_{0}^{(\pm)}\partial_{N}\Phi_{1} + C_{1}^{(\pm)} \partial_{N}\Phi_{0})  =  \nonumber \hspace{55mm} \\
     \frac{1}{D^{(\pm)}} \int_{0}^{N}
        \left(\partial_{t} +\boldsymbol{v}_{s0}\cdot\nabla_{s} + 
                                                         \partial_{n}v_{p}|_{S}N^{\prime} \partial_{N^{\prime}} \right) C_{0}^{(\pm)} dN^{\prime}  .
\label{intNP2drop} 
\end{eqnarray}
This expresses a balance between the electrodiffusive flux of ions in the normal direction and advection of ions within the layers in integrated form.   

To form the limit of (\ref{intNP2drop}) as $N\rightarrow\pm\infty$ we match with the outer regions, which requires that: 
$\partial_{N}C_{1}^{(\pm)}\rightarrow \partial_{n}c_{0}|_{S_{\pm}}$ and $\partial_{N}\Phi_{1}^{(\pm)}\rightarrow \partial_{n}\phi_{0}|_{S_{\pm}}$;  
both $C_{0}^{(\pm)}\rightarrow c_{0}|_{S_{\pm}}$ and $\Phi_{0}\rightarrow\phi_{0}|_{S_{\pm}}$, where the approach to the limit is exponential in $N$ 
from (\ref{C0ep}) and (\ref{GC}).  The limit of the left hand side is therefore $\partial_{n}c_{0}|_{S_{\pm}}\pm c_{0}|_{S_{\pm}}\partial_{n}\phi_{0}|_{S_{\pm}}$, 
in which the upper (lower) choice of the sign that sits between the two groups of terms refers to the positive (negative) ions respectively, so that it is now 
written as $\partial_{n}c_{0}|_{S_{\pm}} (\pm) c_{0}|_{S_{\pm}}\partial_{n}\phi_{0}|_{S_{\pm}}$.  

To see that the integral on the right hand side of (\ref{intNP2drop}) converges as $N\rightarrow\pm\infty$ and to evaluate it in terms of the layer 
$\zeta$-potentials etc., we recall from (\ref{NP1}) and \S \ref{sec:outer} that in the outer regions, leading order ion conservation requires that 
$(\partial_{t}+\boldsymbol{u}_{0}\cdot\nabla)c_{0}=0$ for all $n$, including the limit $n\rightarrow 0^{\pm}$ in which 
$c_{0}=c_{0}|_{S_{\pm}}$.  In the local intrinsic coordinate system, from (\ref{DDt_0}) this becomes  
\begin{equation}
\left( \partial_{t} +\boldsymbol{v}_{s0}\cdot\nabla_{s} \right) c_{0}|_{S_{\pm}}=0 \, ,
\label{dtc0}
\end{equation}
since $c_{0}|_{S_{\pm}}$ is independent of $N$.  When this is subtracted from the integrand of (\ref{intNP2drop}), in the limit we have   
\begin{eqnarray}
\partial_{n}c_{0}|_{S_{\pm}} (\pm) c_{0}|_{S_{\pm}}\partial_{n}\phi_{0}|_{S_{\pm}} = \hspace{72mm} \nonumber \\
\frac{1}{D^{(\pm)}}
\left(  (\partial_{t} +\boldsymbol{v}_{s0}\cdot\nabla_{s})  \int_{0}^{\pm\infty} (C_{0}^{(\pm)}-c_{0}|_{S_{\pm}}) \, dN \right.  \hspace{8mm}  \nonumber \\
\left. + \, \partial_{n}v_{p}|_{S} \int_{0}^{\pm\infty} N\partial_{N} C_{0}^{(\pm)} dN   \right) .
\label{intNP3drop}
\end{eqnarray} 

The last integral on the right hand side of (\ref{intNP3drop}) can be found by integration by parts, where because of the exponential approach 
of $C_{0}^{(\pm)}$ to $c_{0}|_{S_{\pm}}$ as $N\rightarrow \pm \infty$ the antiderivative of $\partial_{N}C_{0}^{(\pm)}$ is chosen to be 
$C_{0}^{(\pm)}-c_{0}|_{S_{\pm}}$, so that 
$\int_{0}^{\pm\infty}N\partial_{N} C_{0}^{(\pm)} dN = - \int_{0}^{\pm\infty} (C_{0}^{(\pm)}-c_{0}|_{S_{\pm}}) dN$. 
In physical terms both integrals in (\ref{intNP3drop}) are now, up to a sign, the excess ion concentration integrated across a layer.  Since the 
excess potential $\eta_{0}$ is a monotone function of $N$ and of the same sign as the $\zeta$-potential, the variable of integration can be changed 
from $N$ to $\eta_{0}$, from which equations (\ref{C0ep}) and (\ref{dNPhi0}) give 
\begin{equation}
\int_{0}^{\pm\infty} (C_{0}^{(\pm)}-c_{0}|_{S_{\pm}}) dN = \pm 2 \sqrt{\varepsilon c_{0}|_{S_{\pm}}} \left( e^{(\mp)\zeta_{0\pm}/2} -1 \right) \, . 
\label{int1drop}
\end{equation}

When equation (\ref{intNP3drop}) is written out for each ion species separately, their sum and difference is formed and (\ref{int1drop}) is used, we 
find a pair of boundary conditions to be applied at the outer edge of each Debye layer.  To do so, recall that in (\ref{intNP3drop}) and (\ref{int1drop}) when 
a choice of sign appears in parenthesis, as in $(\pm)$ and $(\mp)$, the upper (lower) choice of sign refers to the positive (negative) ion species respectively.  
When a choice of sign appears without parenthesis, the upper (lower) choice of sign refers to the Debye layer in the exterior (interior) phase, where $N>0$ 
($N<0$).  This gives 
\begin{equation}
\partial_{n}c_{0}|_{S_{\pm}} =
\pm   \left( \partial_{t} +\boldsymbol{v}_{s0}\cdot\nabla_{s} - \partial_{n} v_{p}|_{S} \right) \sqrt{\varepsilon c_{0}|_{S_{\pm}}} 
\left( \frac{e^{\frac{-\zeta_{0\pm}}{2}}-1}{D^{(+)}}  + \frac{e^{\frac{\zeta_{0\pm}}{2}}-1}{D^{(-)}}\right) , 
\label{Bccn0drop} 
\end {equation}
and 
\begin{equation}
c_{0}|_{S_{\pm}}\partial_{n}\phi_{0}|_{S_{\pm}} = 
\pm \left(\partial_{t} +\boldsymbol{v}_{s0}\cdot\nabla_{s} - \partial_{n} v_{p}|_{S} \right)) \sqrt{\varepsilon c_{0}|_{S_{\pm}}}  
\left( \frac{e^{\frac{-\zeta_{0\pm}}{2}}-1}{D^{(+)}}  - \frac{e^{\frac{\zeta_{0\pm}}{2}}-1}{D^{(-)}}\right) .
\label{Bcphidrop} 
\end{equation}

To conclude this section we note that for a drop the permittivity $\varepsilon$ and ion diffusivities $D^{(\pm)}$ may differ between the interior and exterior 
phases.  When this occurs, the material properties are ascribed a $2$-subscript (or $1$-subscript) when relations (\ref{Bccn0drop}) and (\ref{Bcphidrop}) 
are applied as boundary conditions on $S_{+}$ (or $S_{-}$) for the exterior (or interior) phase.

\subsection{Debye layer ion transport for a vesicle}
\label{sec:ITvesicle}

When compared to a drop, some steps of the analysis for ion transport in the Debye layers need to be modified for a vesicle, since its membrane can be 
semi-permeable to ion gating and solvent osmosis channels.  

First, the expression (\ref{NP2}) for ion conservation in the Debye layers can be written in terms of the normal ion flux 
$\boldsymbol{J}_{1}^{(\pm)}\cdot\boldsymbol{n}$ as 
\begin{equation}
\partial_{N} ( - \boldsymbol{J}_{1}^{(\pm)}\cdot\boldsymbol{n} ) = 
     \left( \partial_{t} +\boldsymbol{v}_{s0}\cdot\nabla_{s} + (v_{p}|_{S} + \partial_{n}v_{p}|_{S}N) \partial_{N} \right) C_{0}^{(\pm)} \, . 
\label{NP3}
\end{equation}
This includes the trans-membrane osmotic flow velocity $v_{p}|_{S}$ in the ion advection terms on the right hand side.  

Next, the trans-membrane ion flux $\boldsymbol{J}_{1}^{(\pm)}\cdot\boldsymbol{n}$ evaluated on $S$, i.e., when $N=0$, is denoted by   
$j|_{S}^{(\pm)}$ and given in reduced form by (\ref{jflux}).  An integration of (\ref{NP3}) with respect to $N$, with the ion flux restored to terms 
of the ion concentrations and potential, gives 
\begin{eqnarray}
\partial_{N}C_{1}^{(\pm)} \pm (C_{0}^{(\pm)}\partial_{N}\Phi_{1} + C_{1}^{(\pm)} \partial_{N}\Phi_{0})  =   \nonumber \hspace{52mm} \\
  \frac{-j|_{S}^{(\pm)}}{D^{(\pm)}}  +  \frac{1}{D^{(\pm)}} \int_{0}^{N}
        \left(\partial_{t} +\boldsymbol{v}_{s0}\cdot\nabla_{s} + 
                                                         (v_{p}|_{S} + \partial_{n}v_{p}|_{S}N^{\prime}) \partial_{N^{\prime}} \right) C_{0}^{(\pm)} dN^{\prime} \, .
\label{intNP2ves} 
\end{eqnarray}

The procedure of matching with the outer regions and subtracting equation (\ref{dtc0}) from the integrand remains unchanged, with the result that 
the analog for a vesicle of (\ref{intNP3drop}) is 
\begin{eqnarray}
\partial_{n}c_{0}|_{S_{\pm}} (\pm) c_{0}|_{S_{\pm}}\partial_{n}\phi_{0}|_{S_{\pm}} = \hspace{72mm} \nonumber \\
  \frac{-j|_{S}^{(\pm)}}{D^{(\pm)}} + \frac{1}{D^{(\pm)}}
\left(  (\partial_{t} +\boldsymbol{v}_{s0}\cdot\nabla_{s})  \int_{0}^{\pm\infty} (C_{0}^{(\pm)}-c_{0}|_{S_{\pm}}) \, dN \right.  \hspace{8mm}  \nonumber \\
\left. +  \int_{0}^{\pm\infty} ( v_{p}|_{S} + \partial_{n}v_{p}|_{S} N ) \partial_{N} C_{0}^{(\pm)} dN   \right) .
\label{intNP3ves}
\end{eqnarray} 
Since the osmotic flow velocity $v_{p}|_{S}$ is independent of $N$, the additional integral term is simply the change in the ion concentrations across 
either Debye layer, which is given by (\ref{C0ep}) and (\ref{zetapsV}) as $c_{0}|_{S_{\pm}}(1-e^{(\mp)\zeta_{0\pm}})$.

When equation (\ref{intNP3ves}) is written for each ion species, and then their sum and difference is formed, we find that for a vesicle the boundary 
conditions (\ref{Bccn0drop}) and (\ref{Bcphidrop}) are modified to become 
\begin{eqnarray}
\partial_{n}c_{0}|_{S_{\pm}} = 
             -  \frac{1}{2} \left( \frac{j|_{S}^{(+)}}{D^{(+)}} + \frac{j|_{S}^{(-)}}{D^{(-)}}  \right) + \frac{v_{p}|_{S} \, c_{0}|_{S_{\pm}}}{2} 
             \left(  \frac{1-e^{-\zeta_{0\pm}}}{D^{(+)}} + \frac{1-e^{\zeta_{0\pm}}}{D^{(-)}}  \right)  \hspace*{15mm} \nonumber \\
\pm  \left( \partial_{t} +\boldsymbol{v}_{s0}\cdot\nabla_{s} - \partial_{n} v_{p}|_{S} \right) \sqrt{\varepsilon c_{0}|_{S_{\pm}}} 
\left( \frac{e^{\frac{-\zeta_{0\pm}}{2}}-1}{D^{(+)}}  + \frac{e^{\frac{\zeta_{0\pm}}{2}}-1}{D^{(-)}}\right) , \hspace*{6mm}
\label{Bccn0ves} 
\end {eqnarray}
and 
\begin{eqnarray}
c_{0}|_{S_{\pm}}\partial_{n}\phi_{0}|_{S_{\pm}} = 
             -  \frac{1}{2} \left( \frac{j|_{S}^{(+)}}{D^{(+)}} - \frac{j|_{S}^{(-)}}{D^{(-)}} \right) + \frac{v_{p}|_{S} \, c_{0}|_{S_{\pm}}}{2}  
             \left(  \frac{1-e^{-\zeta_{0\pm}}}{D^{(+)}} - \frac{1-e^{\zeta_{0\pm}}}{D^{(-)}}  \right)   \hspace*{12mm} \nonumber \\
\pm \left(\partial_{t} +\boldsymbol{v}_{s0}\cdot\nabla_{s} - \partial_{n} v_{p}|_{S} \right)) \sqrt{\varepsilon c_{0}|_{S_{\pm}}}  
\left( \frac{e^{\frac{-\zeta_{0\pm}}{2}}-1}{D^{(+)}}  - \frac{e^{\frac{\zeta_{0\pm}}{2}}-1}{D^{(-)}}\right) , \hspace*{8mm}
\label{Bcphives} 
\end{eqnarray}
where the upper (lower) choice of sign refers to the outer edge of the Debye layer in the exterior (interior) phase, where $N>0$ ($N<0$).  

Expressions for the additional trans-membrane ion flux contributions to (\ref{Bccn0ves}) and (\ref{Bcphives}), written in reduced form 
via (\ref{jflux}), are found on noting that 
\begin{equation}
\frac{j|_{S}^{(+)}}{D^{(+)}} \pm \frac{j|_{S}^{(-)}}{D^{(-)}} = - \left( \frac{g^{(+)}}{D^{(+)}} \pm \frac{g^{(-)}}{D^{(-)}} \right) [ \ln c_{0} ]^{S_{+}}_{S_{-}} 
   - \left( \frac{g^{(+)}}{D^{(+)}} \mp \frac{g^{(-)}}{D^{(-)}} \right) [ \phi_{0} ]^{S_{+}}_{S_{-}} \, .
\label{jfluxphc}
\end{equation}
Similarly, for the osmotic flow contribution we recall from (\ref{vps}) that $v_{p}|_{S} = 2\pi_{m} [ c_{0} ]^{S_{+}}_{S_{-}}$. 

For a vesicle, the same solvent occupies both the exterior ($\Omega_{2}$) and interior ($\Omega_{1}$) phase and we consider single cation and anion 
species, so that material properties such as the permittivity $\varepsilon$ and ion diffusivities $D^{(\pm)}$ are constant throughout.

\subsection{Use of the ion transport relations as boundary conditions}
\label{sec:outer_c}

As noted in \S \ref{sec:outer} the parts of $\Omega_{1}$ and $\Omega_{2}$ in the outer regions, outside the Debye layers, are charge-neutral to high order, 
where the field equations are 
\begin{subeqnarray}
\nabla^{2}\phi_{0} = 0 \, , \hspace{10mm} (\partial_{t}+\boldsymbol{u}_{0}\cdot\nabla)c_{0}=0 \, , \hspace{2mm}  \slabel{outeralla} \\
-\nabla p_{0} +\mu\nabla^{2}\boldsymbol{u}_{0} = 0 \, , \hspace{3mm}   \mbox{and}  \ \ \   \nabla \cdot \boldsymbol{u}_{0} = 0 \, . \slabel{outerallb} 
\label{outerall}
\end{subeqnarray}
Except for the appearance of the fluid velocity $\boldsymbol{u}_{0}$ in the transport equation for $c_{0}$, the field equations are uncoupled.  

The relation (\ref{Bcphidrop}) for a drop or (\ref{Bcphives}) for a vesicle is a Neumann boundary condition for the potential $\phi_{0}$ of the Laplace equation 
that is applied at the outer edge of each Debye layer, that is $S_{+}$ in $\Omega_{2}$ and $S_{-}$ in $\Omega_{1}$.  It depends on the local fluid velocity and 
on the local ion concentration $c_{0}$ and $\zeta$-potential $\zeta_{0}$ at each point of the respective Debye layer edge.  For a vesicle, the boundary condition 
also depends on the jump in both the ion concentration and potential across the double layer pair, per (\ref{jfluxphc}) and the expression for $v_{p}|_{S}$ 
below it. 

The relation (\ref{Bccn0drop}) for a drop or (\ref{Bccn0ves}) for a vesicle is a boundary condition of mixed type for the ion concentration $c_{0}$ of the 
transport equation at (\ref{outeralla}).  It depends on the local fluid velocity and the layer $\zeta$-potential.  For a vesicle it also depends on the jump in the 
ion concentration and on the jump in the potential across the double layer.  Notice that, for a drop, the boundary condition cannot be applied exactly on the interface 
$S$, since the characteristics of the transport equation are particle paths satisfying $\frac{d\boldsymbol{x}}{dt}=\boldsymbol{u}_{0}$ that are tangential to 
$S$ in its Lagrangian frame and therefore do not enter the outer regions.  Instead, for both a drop and a vesicle the boundary condition is applied only on the 
outflow regions of the Debye layers, where the quantity $\partial_{n}v_{p}|_{S}$ introduced in \S \ref{sec:hydro0} is positive, at some fixed $\epsilon>0$ and 
$N\neq 0$.  Further details of this construction will be given elsewhere.

\section{The integral equation for the fluid velocity on the interface}
\label{sec:uint}

In this section a boundary integral equation is formulated for the fluid velocity on the interface.  The subscript $i$ is appended to the material parameters 
$\mu_{i}$ and $\varepsilon_{i}$ in the disjoint domains $\Omega_{i}$, where $i=1, 2$.  A brief derivation is given first for Stokes flow with a general body 
force $\boldsymbol{f}$, so that  
\begin{equation}
-\nabla p + \mu_{i}\nabla^{2}\boldsymbol{u} + \boldsymbol{f} = 0 \ \ \mbox{and} \ \ 
\nabla \cdot \boldsymbol{u}=0. 
\label{Stokesa}
\end{equation}
We then find the form this takes for the problem at hand with $\boldsymbol{f}= \varepsilon_{i} \nu \epsilon \nabla^{2}\phi\nabla\phi$ in the small-$\epsilon$ limit, when the 
body force is confined to a neighborhood of the interface $S$. 

The free-space dyadic Green's function is the solution of 
\begin{equation}
-\nabla p^{\prime} + \mu_{i}\nabla^{2}\boldsymbol{u}^{\prime} + \boldsymbol{g}\delta(\boldsymbol{x}-\boldsymbol{x}_{0}) = 0 \ \ \mbox{and} \ \ 
\nabla \cdot \boldsymbol{u}^{\prime}=0, 
\label{Stokesb}
\end{equation}
where $\boldsymbol{g}$ is an arbitrary constant.  In suffix notation the solution is 
\begin{equation}
u^{\prime}_{i}(\boldsymbol{x})=\frac{1}{8\pi\mu_{(i)}}G_{ij}(\boldsymbol{x}, \boldsymbol{x}_{0})g_{j}, \ \ 
T^{\prime}_{Hik}=\frac{1}{8\pi}T_{ijk}(\boldsymbol{x}, \boldsymbol{x}_{0})g_{j}, \ \ 
p^{\prime}=\frac{1}{8\pi}p_{*j}(\boldsymbol{x}, \boldsymbol{x}_{0})g_{j} .
\label{Stokesc}
\end{equation}
We adopt a convention that a subscript on a material parameter is put in parenthesis when suffix notation is used, e.g. on the local viscosity $\mu_{(i)}$ in (\ref{Stokesc}).  
Explicit expressions for the Stokeslet $G_{ij}(\boldsymbol{x}, \boldsymbol{x}_{0})$, stresslet $T_{ijk}(\boldsymbol{x}, \boldsymbol{x}_{0})$, and pressure 
${p}_{*j}(\boldsymbol{x}, \boldsymbol{x}_{0})$ are 
\begin{equation}
G_{ij}(\boldsymbol{x}, \boldsymbol{x}_{0})=\frac{\delta_{ij}}{r}+\frac{\hat{x}_{i}\hat{x}_{j}}{r^{3}}, \ \ 
T_{ijk}(\boldsymbol{x}, \boldsymbol{x}_{0})= -6 \frac{\hat{x}_{i}\hat{x}_{j}\hat{x}_{k}}{r^{5}}, \ \ 
p_{*j}(\boldsymbol{x}, \boldsymbol{x}_{0})=2\frac{\hat{x}_{j}}{r^{3}}, 
\end{equation}
see, for example, \cite{Pozrikidis_book1}.  Here $\hat{\boldsymbol{x}}=\boldsymbol{x}-\boldsymbol{x}_{0}$ and $r=|\hat{\boldsymbol{x}}|$.  

The reciprocal identity for the flow fields of (\ref{Stokesa}) and (\ref{Stokesb}) is 
\[
\frac{\partial}{\partial x_{k}}(u^{\prime}_{i}T_{Hik}-u_{i}T^{\prime}_{Hik})= u_{i}g_{i}\delta(\boldsymbol{x}-\boldsymbol{x}_{0})-u^{\prime}_{i}f_{i} \, .
\]
Written in terms of the Stokeslet and stresslet, since $\boldsymbol{g}$ is arbitrary, this is 
\begin{eqnarray}
\frac{\partial}{\partial x_{k}}\left(G_{ij}(\boldsymbol{x}, \boldsymbol{x}_{0})T_{Hik}(\boldsymbol{x})-\mu_{(i)}u_{i}(\boldsymbol{x})T_{ijk}(\boldsymbol{x}, \boldsymbol{x}_{0})\right) 
                        =  \hspace{35mm} \nonumber \\
 \hspace{25mm}  8\pi \mu_{(i)} u_{j}(\boldsymbol{x})\delta(\boldsymbol{x}-\boldsymbol{x}_{0})-G_{ij}(\boldsymbol{x}, \boldsymbol{x}_{0})f_{i}(\boldsymbol{x}) \, .
\label{recip}
\end{eqnarray}
The procedure of fixing the location of the pole or target point $\boldsymbol{x}_{0}$, and here we choose $\boldsymbol{x}_{0}\in\Omega_{2}$, then integrating over 
$\boldsymbol{x}\in \Omega_{1}$ and over $\boldsymbol{x}\in \Omega_{2}$ in turn, applying the divergence theorem, and noting that the stress tensor 
$\boldsymbol{T}_{H}(\boldsymbol{x}, \boldsymbol{x}_{0})$ and body force $\boldsymbol{f}(\boldsymbol{x})$ may have finite jump discontinuities across the interface $S$, while 
$\boldsymbol{u}(\boldsymbol{x})$ is continuous there, gives the integral representation
\begin{eqnarray}
u_{j}(\boldsymbol{x}_{0})=-\frac{1}{8\pi\mu_{2}}\int_{S}G_{ij}(\boldsymbol{x}, \boldsymbol{x}_{0})[T_{Hik}(\boldsymbol{x})]n_{k}(\boldsymbol{x}) dS_{\boldsymbol{x}}  
           \hspace{47mm}  \nonumber \\
+\frac{\mu_{2}-\mu_{1}}{8\pi\mu_{2}} \int_{S} u_{i}(\boldsymbol{x}) T_{ijk}(\boldsymbol{x}, \boldsymbol{x}_{0}) n_{k}(\boldsymbol{x}) dS_{\boldsymbol{x}}  + 
\frac{1}{8\pi\mu_{2}}\int_{\Omega_{1}\cup\Omega_{2}}G_{ij}(\boldsymbol{x}, \boldsymbol{x}_{0})f_{i}(\boldsymbol{x}) dV_{\boldsymbol{x}} \, , \hspace{2mm}
\label{uinteqn_off}
\end{eqnarray}
for the fluid velocity $\boldsymbol{u}$ at $\boldsymbol{x}_{0}\in\Omega_{2}$.

In the limit when $\boldsymbol{x}_{0}$ approaches $S$ from $\Omega_{2}$ the stresslet integral of (\ref{uinteqn_off}) has an order one local contribution 
from a neighborhood of $\boldsymbol{x}_{0}$, which is $4\pi u_{j}(\boldsymbol{x}_{0})$.  This leads to the boundary integral equation 
\begin{eqnarray}
u_{j}(\boldsymbol{x}_{0}) = \frac{-1}{4\pi(\mu_{2}+\mu_{1})}\int_{S}G_{ij}(\boldsymbol{x}, \boldsymbol{x}_{0})[T_{Hik}(\boldsymbol{x})]n_{k}(\boldsymbol{x}) dS_{\boldsymbol{x}}  
           \hspace{25mm}  \nonumber \\
+\frac{(\mu_{2}-\mu_{1})}{4\pi(\mu_{2}+\mu_{1})} \int^{PV}_{S} u_{i}(\boldsymbol{x}) T_{ijk}(\boldsymbol{x}, \boldsymbol{x}_{0}) n_{k}(\boldsymbol{x}) dS_{\boldsymbol{x}}  
        \hspace{15mm}  \nonumber \\  
      + \frac{1}{4\pi(\mu_{2}+\mu_{1})}\int_{\Omega_{1}\cup\Omega_{2}}G_{ij}(\boldsymbol{x}, \boldsymbol{x}_{0})f_{i}(\boldsymbol{x}) dV_{\boldsymbol{x}} \, ,  \hspace{5mm} 
\label{uinteqn}
\end{eqnarray}
for the fluid velocity $\boldsymbol{u}$ at $\boldsymbol{x}_{0}\in S$, where $PV$ denotes the principal value of the improper integral.  This holds for both a drop and a vesicle.

The integral equation (\ref{uinteqn}) can also be constructed by placing the target point $\boldsymbol{x}_{0}$ at an arbitrary location on the interface $S$ at the outset, 
and excising a small sphere $B_{\delta}(\boldsymbol{x}_{0})$ centered on $\boldsymbol{x}_{0}$ with radius $\delta$ from the region of integration.  This removes the 
contribution of the $\delta$-function term in the reciprocal identity (\ref{recip}).  With $B_{\delta}(\boldsymbol{x}_{0})$ removed, the reciprocal identity is integrated over the 
interior $\boldsymbol{x}\in \Omega_{1}$, and then over that part of the exterior $\boldsymbol{x}\in \Omega_{2}$ bounded by a large sphere $B_{R}$ of radius 
$R$.  After the divergence theorem is applied, the contribution from the surface integral over $B_{R}$ vanishes in the limit $R\rightarrow\infty$, and in the limit 
$\delta\rightarrow 0$ integration of the stresslet term over the semi-spherical surface of $B_{\delta}(\boldsymbol{x}_{0})$ makes a local contribution  
$4\pi\mu_{(i)}u_{j}(\boldsymbol{x}_{0})$ in each domain $\Omega_{i}$, when the unit normal points out from $\boldsymbol{x}_{0}$.  See, for example, \cite{Rallison&Acrivos}.

\subsection{The net force on the interface and the integral equation for a drop}
\label{sec:forces_drop}
 
The traction on the interface $S$ due to the hydrodynamic field appears in the first integral on the right hand side of (\ref{uinteqn}) and can be expressed in terms of 
the capillary stress and electrostatic potential from the stress-balance boundary condition for a drop (\ref{sbal}) and the Maxwell stess tensor (\ref{tensnondim}).  This 
gives the exact relation 
\begin{equation}
[\boldsymbol{T}_{H}\cdot\boldsymbol{n}] = \Delta (\kappa_{1}+\kappa_{2})\boldsymbol{n} - \nu\epsilon\, [\, \varepsilon_{i}(\frac{1}{2}(\partial_{n}\phi^{2}-| \nabla_{s}\phi |^{2})\boldsymbol{n} + 
\partial_{n}\phi\nabla_{s}\phi)] \, .
\label{hydrostressS}
\end{equation}
Since the drop interface is sharp, with no surface charge and no electrical capacitance, the electrostatic potential $\phi$ and its tangential 
gradient $\nabla_{s}\phi$ are continuous across $S$, and, per the second of equations (\ref{BC1a}), $[\varepsilon\partial_{n}\phi]=0$.  On the right hand side of 
(\ref{hydrostressS}), the last term is therefore zero.  Recalling the $\epsilon$-scaling and local expansion of $\phi$ in the Debye layers, we find that the preceding term, 
containing $|\nabla_{s}\phi|^{2}$, is of order $O(\epsilon)$, so that the hydrodynamic traction on $S$ is 
\begin{equation}
[\boldsymbol{T}_{H}\cdot\boldsymbol{n}] =  \Delta (\kappa_{1}+\kappa_{2})\boldsymbol{n} -  
\frac{\nu}{2\epsilon} \, [\, \varepsilon_{i} (\partial_{N}\Phi_{0})^{2}\,]  \boldsymbol{n}  
 -\nu \, [\, \varepsilon_{i} \partial_{N}\Phi_{0}\partial_{N}\Phi_{1}] \,  \boldsymbol{n} + O(\epsilon) \, .  
\label{hydrofinal}
\end{equation}

It turns out that the last two terms on the right hand side of (\ref{hydrofinal}) are cancelled by contributions to the total force on the electric charge distribution 
within the layers, which is given by the last integral on the right hand side of (\ref{uinteqn}) containing the Coulomb force density $\boldsymbol{f}$.  This integral is 
recast as a surface integral by performing the integration in the normal direction over the $\epsilon$-support of the Debye layers, for which we give details in 
Appendix \ref{sec:appA}.  

The result of this analysis is that the integral equation (\ref{uinteqn}) can be written as 
\begin{eqnarray}
u_{j}(\boldsymbol{x}_{0}) = \frac{-1}{4\pi(\mu_{2}+\mu_{1})}\int_{S}G_{ij}(\boldsymbol{x}, \boldsymbol{x}_{0})[\hat{T}_{ik}(\boldsymbol{x})]n_{k}(\boldsymbol{x}) dS_{\boldsymbol{x}}  
           \hspace{32mm}  \nonumber \\
+\frac{(\mu_{2}-\mu_{1})}{4\pi(\mu_{2}+\mu_{1})} \int^{PV}_{S} u_{i}(\boldsymbol{x}) T_{ijk}(\boldsymbol{x}, \boldsymbol{x}_{0}) n_{k}(\boldsymbol{x}) dS_{\boldsymbol{x}}  
        \hspace{19mm}  \nonumber \\  
      - \frac{1}{4\pi(\mu_{2}+\mu_{1})}\int_{S} n_{k}(\boldsymbol{x})\frac{\partial}{\partial x_{k}} G_{ij}(\boldsymbol{x}, \boldsymbol{x}_{0}) \nu \Theta(\boldsymbol{x}) n_{i}(\boldsymbol{x}) dS_{\boldsymbol{x}}  \, .
                            \hspace{4mm}
\label{uinteqn2}
\end{eqnarray}
Here, $[\boldsymbol{\hat{T}}\cdot\boldsymbol{n}]$ is a modified surface traction, given by 
\begin{equation}
[\boldsymbol{\hat{T}}\cdot\boldsymbol{n}] = (\kappa_{1}+\kappa_{2})\left( \Delta - \nu\Theta \right)\boldsymbol{n} + \nu \nabla_{s}\Theta \, ,
\label{tracnetD}
\end{equation}
and where $\Theta$ is given in terms of the layer $\zeta$-potentials and other parameters by  
\begin{equation}
\Theta= 4 \left( \surd({\varepsilon_{2}c_{0}|_{S_{+}}}) \sinh^{2}\frac{\zeta_{0+}}{4} +   \surd({\varepsilon_{1}c_{0}|_{S_{-}}}) \sinh^{2}\frac{\zeta_{0-}}{4} \right) \, .
\label{Thetadef}
\end{equation}

The quantity $\Theta$ is defined in Appendix \ref{sec:appA} at equation (\ref{10}) by $\Theta=\int_{-\infty}^{\infty}\varepsilon \frac{\partial_{N}\Phi_{0}^{2}}{2} dN$,  
which brings to mind the electrostatic energy density per unit area in the Debye layers.  To verify this conjecture, at leading order, the dimensions can be 
restored to equations (\ref{uinteqn2}) to (\ref{Thetadef}), when it is found that they still hold with the only modifications that: $\nu$ is set to 1 in equations (\ref{uinteqn2}) 
and (\ref{tracnetD}), $\Delta$ is replaced by the surface tension $\sigma$ in (\ref{tracnetD}), and the $\zeta$-potentials in (\ref{Thetadef}) are replaced by 
$\zeta_{0\pm}/\phi_{*}$ so as to remain dimensionless.  Then the scale by which $\Theta$ is made dimensionless is found to be $\varepsilon_{*}\phi_{*}^{2}/\lambda_{*}$, 
which is consistent with the $O(1/\epsilon)$ magnitude of the potential gradient and the $\epsilon$-width of the Debye layers.

\subsection{Modification of the net force on the interface and the integral equation for a vesicle}
\label{sec:forces_vesicle}

Relative to a drop with constant uniform surface tension, the stress in a vesicle membrane includes the effects of bending stiffness and a surface tension $\sigma$ that 
varies so as to maintain local conservation of area via the inextensibility constraint $\nabla_{s}\cdot(\partial_{t}\boldsymbol{X}_{s}(\boldsymbol{\theta}, t))=0$ of (\ref{inext}).  
This modifies the stress-balance boundary condition (\ref{hydrostressS}) by replacing the capillary stress $\Delta (\kappa_{1}+\kappa_{2})\boldsymbol{n}$ on the right hand 
side of the relation with the membrane stress
\begin{equation}
\boldsymbol{\tau}_{m} =  2 \sigma H \boldsymbol{n} - \nabla_{s} \sigma  -  \kappa_{b} \left( 2H(H^{2}-K)  +  \nabla_{s}^{2} H \right) \boldsymbol{n} \, , 
\label{memstress}
\end{equation}
per (\ref{sbal}) and (\ref{cellsbal}). 

The second modification to the analysis for a vesicle membrane concerns the jump in the tangential Maxwell stress, which is proportional to 
$[\varepsilon_{i}\partial_{n}\phi\nabla_{s}\phi]$ in (\ref{hydrostressS}).  For a membrane that has no net monopole charge the electric displacement is 
continuous, so that $[\varepsilon_{i}\partial_{n}\phi]=0$ per the boundary condition (\ref{capbc}).  However, the trans-membrane potential $[\phi]$ is nonzero and varies 
around the membrane surface $S$, so that $[\nabla_{s}\phi]\neq 0$, and the tangential Maxwell stress jump is therefore also nonzero.  However, the electrostatic force 
in the Debye layers has a component that, up to the order of calculation, exactly cancels this contribution to the interfacial traction.  This point is made in the analysis 
of Appendix \ref{sec:appA}, below equation (\ref{14}) at item (i).  

It follows that the integral equation (\ref{uinteqn2}) also holds for a vesicle, but with the surface traction (\ref{tracnetD}) replaced by 
\begin{equation}
[\boldsymbol{\hat{T}}\cdot\boldsymbol{n}] = \left\{ (\kappa_{1}+\kappa_{2}) (\sigma - \nu\Theta) - \kappa_{b} (2H(H^{2}-K) + \nabla_{s}^{2} H) \right\} \boldsymbol{n} 
            - \nabla_{s} (\sigma - \nu\Theta) \, .
\label{tracnetV}
\end{equation}
The Stokes dipole integral remains unchanged.  However, as noted below equation (\ref{jfluxphc}) for example, since the same solvent and ion species occupy both 
phases, the permittivity is either the same constant throughout or nearly so, so that in the expression (\ref{Thetadef}) for the energy density $\Theta$, 
$\varepsilon_{1}=\varepsilon_{2}$.

\section{The integral equations for the electrostatic potential on either side of the interface}
\label{sec:phi_int}

In the charge-neutral outer regions away from the interface, the electrostatic potential $\phi=\phi_{0}+\epsilon \phi_{1}+\ldots$ is such that $\phi_{0}$ 
satisfies Laplace's equation, as noted in \S \ref{sec:outer}.  A fundamental result of potential theory gives the potential $\phi_{0}$ on either side of $S$ 
in terms of its normal derivative $\partial_{n}\phi_{0}$ via solution of a Fredholm second type integral equation \cite{Kress}, where the data for the normal 
derivative is considered to be known from equation (\ref{Bcphidrop}) for a drop or (\ref{Bcphives}) for a vesicle. 

With the normal directed outward from $\Omega_{1}$ to $\Omega_{2}$, the potential $\phi_{0}|_{S_{-}}$ on the Debye layer edge $S_{-}$ interior to the interface 
$S$ satisfies,  
\begin{equation}
\int_{S}^{PV}\phi_{0}|_{S_{-}}(\boldsymbol{x})\, \frac{\partial G}{\partial n}(\boldsymbol{x}, \boldsymbol{x_{0}}) \, dS(\boldsymbol{x}) - 
\frac{1}{2} \phi_{0}|_{S_{-}}(\boldsymbol{x_{0}}) = 
\int_{S} G(\boldsymbol{x}, \boldsymbol{x_{0}}) \left. \frac{\partial \phi_{0}}{\partial n}\right|_{S_{-}} \! \! \! \! \! \! \! (\boldsymbol{x}) \, dS (\boldsymbol{x}) \, .
\label{intphiin}
\end{equation}
The potential $\phi_{0}|_{S_{+}}$ on the Debye layer edge $S_{+}$ exterior to the interface satisfies, 
\begin{eqnarray} 
\hspace{5mm}  \int_{S}^{PV}\phi_{0}|_{S_{+}}(\boldsymbol{x})\, \frac{\partial G}{\partial n}(\boldsymbol{x}, \boldsymbol{x_{0}})\, dS(\boldsymbol{x}) +  
\frac{1}{2} \phi_{0}|_{S_{+}}(\boldsymbol{x_{0}}) &=& 
\int_{S} G(\boldsymbol{x}, \boldsymbol{x_{0}}) \left. \frac{\partial \phi_{0}}{\partial n}\right|_{S_{+}} \! \! \! \! \! \! \! (\boldsymbol{x}) \, dS (\boldsymbol{x}) 
                                              \nonumber \hspace{12mm} \\
            & & \hspace{10mm}  + w(\phi_{\infty}, \boldsymbol{x_{0}}) \, ,  
\label{phi+one} \\  
\mbox{where} \hspace{63mm} & &  \nonumber \\
w(\phi_{\infty}, \boldsymbol{x_{0}}) = 
\int_{S}^{PV} \phi_{\infty} (\boldsymbol{x}) \frac{\partial G}{\partial n}(\boldsymbol{x}, \boldsymbol{x_{0}}) \, dS (\boldsymbol{x}) &-& 
\int_{S} G(\boldsymbol{x}, \boldsymbol{x_{0}}) ) \frac{\partial \phi_{\infty}}{\partial n}(\boldsymbol{x}) \, dS (\boldsymbol{x}) \nonumber \\ 
          &  & \hspace{10mm} + \frac{1}{2}\phi_{\infty}(\boldsymbol{x_{0}}) \, .
\end{eqnarray}
Here the free-space Green's function is $G(\boldsymbol{x}, \boldsymbol{x_{0}})=-1/(4\pi |\boldsymbol{x}-\boldsymbol{x_{0}}|)$, and for a leading 
order model the integration is over the interface $S$.

\section{Model summary}
\label{sec:summary}

In this section we give a summary of the model by collecting together results from the text or by indicating where they may be found, or both, 
begining with the more basic components.   Since it is a closed leading order model, all $0$-subscripts are now omitted or understood as being omitted. 

{\it \underline{Initial Conditions.}} The initial conditions of (\ref{ics}), for which the applied field is uniform, are 
\begin{equation} 
\phi(\boldsymbol{x}, 0^{+}) = - \Psi z, \ \boldsymbol{u}(\boldsymbol{x}, 0) = 0, \ \mbox{and} \ c^{(\pm)}(\boldsymbol{x}, 0) = c_{i} \ \mbox{on} \ \Omega_{i}, \ i=1,2\, . 
\label{ics-2}
\end{equation}
An exception to this is a drop that has a nonzero Galvani potential.  Then, for $t<0$ the potential is piecewise constant with jump across $S$ given by 
(\ref{gp2}), namely 
\begin{equation}
[\phi]^{S_{+}}_{S_{-}} =\ \frac{1}{2} \ln \left(\frac{l^{(-)}}{l^{(+)}}\right) \, .
\label{gp2-2}
\end{equation}
The potential at $t=0^{+}$ is given by superimposing the applied potential.  The ambient ion concentrations on $\Omega_{i}$ for $t<0$ are related by (\ref{gp3}), 
that is 
\begin{equation}
c_{1}=c_{2} \sqrt{l^{(+)}l^{(-)}} \, ,
\label{gp3-2}
\end{equation} 
where the ion partition coefficients $l^{(\pm)}$ are known.  The $\zeta$-potentials for $t<0$ that correspond to this are given by (\ref{Galvanizetas}).  A drop is 
initially spherical with radius 1, whereas a vesicle has a known initial equilibrium configuration $S_{ref}$.  

{\it \underline{Far-Field Conditions.}} Far from a drop or vesicle, for a uniform applied field 
\begin{equation}
\phi(\boldsymbol{x}, t) \sim - \Psi z, \  \boldsymbol{u}(\boldsymbol{x}, t) \rightarrow 0, \ \mbox{and} \  
c^{(\pm)}(\boldsymbol{x}, t)\rightarrow \  c_{2} \ \mbox{as} \ |\boldsymbol{x}|\rightarrow \infty \ \mbox{for} \ t > 0 \, . 
\label{far-field-2}
\end{equation}
For a more general applied field $\phi$ approaches a specified solution $\phi_{\infty}$ of Laplace's equation that has magnitude $\Psi$.  This may be time-dependent 
on the scale of the charge-up time.

{\it \underline{Kinematic Condition.}}  Time-update of the interface position is determined by the kinematic condition.  For a drop this is given by 
\begin{equation}
 (\boldsymbol{u} - \frac{\partial\boldsymbol{x}_{s}}{\partial t})\cdot\boldsymbol{n}=0 \ \ \mbox{on}\ S \, ,
\label{kin-2}
\end{equation}
per (\ref{BC2}), at any point  $\boldsymbol{x}_{s}$ on $S$.  The continuity of velocity across $S$, $[\boldsymbol{u}]=0$, is built-in to both the 
integral equation for $\boldsymbol{u}$ and the reduced stress-balance boundary condition.  

For a vesicle with a semi-permeable membrane $S$ having equation $\boldsymbol{x}=\boldsymbol{X}(\boldsymbol{\theta}, t)$, we have the boundary condition 
\begin{equation}
\boldsymbol{u} - \left.\frac{\partial\boldsymbol{X}}{\partial t}\right|_{\boldsymbol{\theta}} = \epsilon v_{p}|_{S} \, \boldsymbol{n} \ \ \ \mbox{on $S$, where} \ \ \ 
v_{p}|_{S} = 2 \pi_{m} [c]^{S_{+}}_{S_{-}} \, ,
\label{solflux-2}
\end{equation}
per (\ref{solfluxmem}) and (\ref{vps}) via (\ref{v_p2b}).  The normal component of this is a slightly modified version of the relation (\ref{kin-2}) for a drop;  when 
the permeability $\pi_{m}>0$ there is a small $O(\epsilon)$ osmotic flux and change in vesicle volume on the charge-up time scale.  The tangential projection of 
(\ref{solflux-2}) implies continuity of fluid velocity across $S$, i.e., $[\boldsymbol{u}]=0$, which is already built-in to other components of the model where needed, 
but in addition it ensures no slip between the membrane and adjacent fluid.  That is, $\boldsymbol{u}_{s}= \partial_{t}\boldsymbol{X}_{s}|_{\boldsymbol{\theta}}$, 
so that local conservation of membrane area or incompressibility implies the additional constraint that 
\begin{equation}
\nabla_{s}\cdot\left( \frac{\partial}{\partial t}\boldsymbol{X}_{s}(\boldsymbol{\theta}, t) \right) =0 
\label{inext-2}
\end{equation}
of (\ref{inext}).  It also ensures that the tangential fluid velocity $\boldsymbol{u}_{s}$ is such that $\nabla_{s}\cdot\boldsymbol{u}_{s}=0$.  

{\it \underline{Inflow-Outflow Condition.}}  The rate of extension of an infinitesimal fluid element based on and normal to $S$ is denoted by $\partial_{n}v_{p}|_{S}$, 
and given in terms of surface data by (\ref{v_p1b}), which is
\begin{equation}
\partial_{n}v_{p}|_{S} = - (\nabla_{s} \cdot \boldsymbol{u}_{s}+(\kappa_{1}+\kappa_{2})u_{n}) \, .
\label{v_p1b-2d}
\end{equation}
For a vesicle with an incompressible membrane, from the comment below equation (\ref{inext-2}), this simplifies to  
\begin{equation}
\partial_{n}v_{p}|_{S} = - (\kappa_{1}+\kappa_{2})u_{n} \, .
\label{v_p1b-2b}
\end{equation}

{\it \underline{$\zeta$-Potentials and Potential Jumps.}} The definition of the $\zeta$-potentials for a drop, for which there is no jump in the potential across its 
sharp interface, is given at (\ref{zetapsD}), from which we have 
\begin{equation}
[\phi]^{S_{+}}_{S_{-}} = \zeta_{-}-\zeta_{+} \, .
\label{zetapsD-2}
\end{equation}
For a vesicle, which can sustain a jump in the potential across its membrane, the definition of the $\zeta$-potentials is given at (\ref{zetapsD}), from which
\begin{equation}
[\phi]^{S_{+}}_{S_{-}} -[\phi]_{m} = \zeta_{-}-\zeta_{+} \, .
\label{zetapsV-2}
\end{equation}
In this summary, the jump in a quantity across the faces of the membrane, which earlier was denoted by $[\cdot] = [\cdot]^{\partial\Omega_{2}}_{\partial\Omega_{1}}$, 
is now denoted by $[\cdot]_{m}$ instead. 

{\it \underline{Gauss's Relation Between the $\zeta$-Potentials.}}  The drop interface holds no monopole charge, from which Gauss's law implies the relation 
(\ref{Gaussdrop}) between the $\zeta$-potentials and the ion concentrations at the outer edges of the Debye layers, namely
\begin{equation}
\sqrt{\varepsilon_{2} c|_{S_{+}}}\sinh \frac{\zeta_{+}}{2} = - \sqrt{\varepsilon_{1} c|_{S_{-}}} \sinh \frac{\zeta_{-}}{2} \, . 
\label{Gaussdrop-2}
\end{equation}
This also implies that the charge per unit area in the back-to-back Debye layers is equal and opposite at all points.  For a vesicle, Gauss's law relates the charge 
held in the layers to the membrane capacitance $C_{m}$ and the trans-membrane potential $[\phi]_{m}$, per (\ref{Gausscell}), namely 
\begin{equation}
\sqrt{\varepsilon_{2} c|_{S_{+}}}\sinh \frac{\zeta_{+}}{2} = - \sqrt{\varepsilon_{1} c|_{S_{-}}} \sinh \frac{\zeta_{-}}{2} = 
    - \frac{C_{m}}{2} [\phi]_{m} \, . 
\label{Gausscell-2}
\end{equation}

At this point we see that, relative to a drop, the model for a vesicle contains an additional variable $[\phi]_{m}$ and an additional constraint in (\ref{Gausscell-2}).  

{\it \underline{Transport Relations for the Potential.}}  For a drop, the relation (\ref{Bcphidrop}) for ion transport in the Debye layers gives the boundary condition 
on $S_{+}$ that 
\begin{equation}
c|_{S_{+}}\partial_{n}\phi|_{S_{+}} = 
 \left(\partial_{t} +\boldsymbol{v}_{s}\cdot\nabla_{s} - \partial_{n} v_{p}|_{S} \right)) \sqrt{\varepsilon_{2} c|_{S_{+}}}  
\left( \frac{e^{\frac{-\zeta_{+}}{2}}-1}{D^{(+)}_{2}}  - \frac{e^{\frac{\zeta_{+}}{2}}-1}{D^{(-)}_{2}}\right) .
\label{Bcphidrop-2+} 
\end{equation}
Similarly, on $S_{-}$ we have
\begin{equation}
c|_{S_{-}}\partial_{n}\phi|_{S_{-}} = 
 - \left(\partial_{t} +\boldsymbol{v}_{s}\cdot\nabla_{s} - \partial_{n} v_{p}|_{S} \right)) \sqrt{\varepsilon_{1} c|_{S_{-}}}  
\left( \frac{e^{\frac{-\zeta_{-}}{2}}-1}{D^{(+)}_{1}}  - \frac{e^{\frac{\zeta_{-}}{2}}-1}{D^{(-)}_{1}}\right) .
\label{Bcphidrop-2-} 
\end{equation}

These give the normal derivative $\partial_{n}\phi$ in terms of flow field quantities on $S$ and the $\zeta$-potential and ion concentration at the outer edge of 
each respective Debye layer, $S_{+}$ or $S_{-}$.  This can be used either as data for the integral equations (\ref{intphiin}) and (\ref{phi+one}) for $\phi$ or for the 
construction of a solution to Laplace's equation 
\begin{equation}
\nabla^{2}\phi = 0
\label{Laplace-2}
\end{equation}
by some other means.  

For a vesicle, analogous data for $\partial_{n}\phi$ in the exterior and interior phase is given by (\ref{Bcphives}), which includes a contribution from the 
trans-membrane ion flux (\ref{jfluxphc}) and osmotic solvent flux (\ref{vps}).  These trans-membrane effects introduce an additional coupling via the jumps in 
the potential $[\phi]^{S_{+}}_{S_{-}}$ and ion concentration $[c]^{S_{+}}_{S_{-}}$ across the outer edges of the double layer pair.

{\it \underline{Transport Relations for the Ion Concentrations.}}  For a drop, the relation (\ref{Bccn0drop}) for Debye layer ion transport gives the boundary condition 
on $S_{+}$ that 
\begin{equation}
\partial_{n}c|_{S_{+}} =
      \left( \partial_{t} +\boldsymbol{v}_{s}\cdot\nabla_{s} - \partial_{n} v_{p}|_{S} \right) \sqrt{\varepsilon_{2} c|_{S_{+}}} 
\left( \frac{e^{\frac{-\zeta_{+}}{2}}-1}{D^{(+)}_{2}}  + \frac{e^{\frac{\zeta_{+}}{2}}-1}{D^{(-)}_{2}}\right) ,
\label{Bccn0drop9-2+} 
\end {equation}
and on $S_{-}$
\begin{equation}
\partial_{n}c|_{S_{-}} =
  -  \left( \partial_{t} +\boldsymbol{v}_{s}\cdot\nabla_{s} - \partial_{n} v_{p}|_{S} \right) \sqrt{\varepsilon_{1} c|_{S_{-}}} 
\left( \frac{e^{\frac{-\zeta_{-}}{2}}-1}{D^{(+)}_{1}}  + \frac{e^{\frac{\zeta_{-}}{2}}-1}{D^{(-)}_{1}}\right) . 
\label{Bccn0drop9-2-} 
\end {equation}

These are boundary conditions for the transport equation 
\begin{equation}
(\partial_{t}+\boldsymbol{u}\cdot\nabla)c=0
\label{transport-2}
\end{equation} 
in both the exterior and interior phase outer regions that, as noted in \S \ref{sec:outer_c}, must be applied away from or off $S$ at some chosen small value of the 
normal coordinate $n \neq 0$.  Further, they can only be applied on regions of outflow from the Debye layers to the bulk, which are such that the quantity 
$\partial_{n}v_{p}|_{S}$ of (\ref{v_p1b-2d}) is positive, i.e., where $\partial_{n}v_{p}|_{S}>0$, since otherwise the problem for the ion transport equation outside the 
Debye layers is over-determined. 

For a vesicle, analogous boundary conditions are given by (\ref{Bccn0ves}), which include the influence of trans-membrane ion and osmotic flux terms given by 
(\ref{jfluxphc}) and (\ref{vps}).  In the outflow condition $\partial_{n}v_{p}|_{S}>0$ that determines where the boundary conditions can be applied, membrane 
incompressibility implies that $\partial_{n}v_{p}|_{S}$ is given by the relation (\ref{v_p1b-2b}).

The boundary conditions for both a drop and a vesicle show coupling from the flow field on $S$ and the layer $\zeta$-potentials to the ion concentration in the outer 
regions away from the Debye layers.  For a vesicle, additional coupling is induced when the trans-membrane fluxes are included. 

{\it \underline{Determination of the flow field.}}  The fluid velocity $\boldsymbol{u}$ on the interface can be found from the integral equation (\ref{uinteqn2}) together 
with its ancillary components.  For a drop these are the net or modified traction (\ref{tracnetD}) and the electrostatic energy density (\ref{Thetadef}).  For 
a vesicle the modified traction is given by (\ref{tracnetV}), which includes the effects of membrane bending stiffness and incompressibility, and in the energy 
density the electrical permittivity is constant throughout the interior and exterior phases.

Instead of solution via an integral equation, the unforced equations of Stokes flow 
\begin{equation}
-\nabla p + \mu_{i}\nabla^{2}\boldsymbol{u} = 0 \ \ \mbox{and} \ \ \nabla \cdot \boldsymbol{u} = 0 \ \ \mbox{on} \ \Omega_{i}, \ \ i=1,2 ,
\label{}
\end{equation}
can be solved by some other means, for example by Lamb's general solution, with the reduced stress-balance boundary conditions for a drop or vesicle that are 
given in Appendix \ref{sec:appB}.  In this formulation, since the model is closed at leading order, the tangential fluid velocity is continuous across the outer edges of 
the double layer pair but there is a jump in pressure and there can be a mismatch in the viscosity $\mu_{1}\neq\mu_{2}$.

Either formulation shows coupling to the flow field on the interface from the electrostatic and ion concentration fields.

\section{Concluding remarks}
\label{sec:conclusion}

We have derived a model for the deformation of either a drop or a vesicle in an applied electric field that is caused by electrokinetic flow or induced 
charge electro-osmosis.  It is derived from the Poisson-Nernst-Planck (PNP) equations for dilute electrolyte solutions coupled to the zero-Reynolds number equations 
of Stokes flow, in the limit where both the interior and exterior phase are strong electrolytes, and the Debye layers that contain the induced charge are thin relative to 
the linear size of the inclusion under equilibrium conditions.  The induced charge is contained in two Debye layers of opposite polarity that form back-to-back on opposite 
sides of the sharp drop interface or vesicle membrane. This configuration is referred to as an electrical double layer, and here it is of liquid-liquid type as opposed to the 
more widely studied solid-liquid type.  

Three distinct fields are present and mutually coupled: the electrostatic field, the hydrodynamic field, and the ion concentration field.  The derivation leads to a 
reduced asymptotic or macroscale model that is closed at leading order, in which the structure of the Debye layers for the three fields is collapsed onto a single surface, 
whose opposite faces $S_{+}$ and $S_{-}$ are the outer edges of the Debye layers at the microscopic scale of the original governing equations. 

The point of view has been to give a derivation that is no more than systematic.  No modeling insight has been applied.  In this respect the evolution of the leaky dielectric model 
has been quite different.   Taylor compared his theoretical predictions on the circulation produced in a
drop by an electric field \cite{Taylor_1966} primarily with the experimental results of Allan and Mason \cite{Allan_Mason1962}, obtaining good agreement based largely on existing 
electrodynamic models for dielectrics.  By the time of Melcher and Taylor's work \cite{Melcher_Taylor_1969} this had developed into the leaky dielectric model of electrohydrodynamics.  
Later, as described in the Introduction, it has been shown that the leaky dielectric model can be underpinned theoretically as the thin Debye layer limit of the Poisson-Nernst-Planck 
equations for weak electrolytes or, equivalently, at low ion densities \cite{Saville_1997,SchnitzerYariv,MoriYoung}. 

It is generally believed that a model that is derived in a particular limit may still provide predictions that are valuable and have some accuracy in regimes and circumstances 
outside the confines in which the model was originally derived.  This could be considered a defining attribute of a good model.  The main difference between the leaky dielectric 
model and that of the present study is our premise of strong electrolytes, and hence relatively high ion densities.  

The similarities or differences that the models predict is, it is hoped, a topic for future work.  But here we note an aspect concerning the polarity of the double layer and the 
direction of its dipole moment.  The component of the local electric field normal to the interface that separates the double layer charge is opposite or anti-parallel to the dipole 
moment of the induced charge.  This is as sketched in Figure \ref{fig:setup}.  Surface tension, bending stiffness, and effects of viscous flow can stabilize the interface but on 
its own this dipole orientation is destabilizing.  It seems possible that this could be a dominant effect at high applied field strengths and induced charge densities.  

The nonlinearity of the Poisson-Boltzmann equation has been kept, in addition to arbitrary curvature, which leads to a mutual coupling between the electrostatic, 
hydrodynamic, and ion concentration fields inside the double layer.  This coupling is simplified at low applied field strengths, when the expressions in the $\zeta$-potentials 
are linearized and the redistribution of ions in the double layer is reduced.  

We have included the vesicle membrane model together with that for a drop.  The vesicle model indicates one way in which a semi-permeable interface can be introduced 
to the analysis as well as incorporating mechanical bending stiffness.  It has been noted elsewhere and in related contexts that an area-preserving or incompressible 
membrane greatly constrains the flow field relative that of a drop.  See, for example, the studies  \cite{Vlahovska_Gracia_2007,Schwalbe_etal_2011,WoodhouseGoldstein2012}. 

\vspace{3mm}

\noindent
{\bf  Acknowledgement} The authors gratefully acknowledge support from National Science Foundation grants DMS-1412789 and DMS-1909407.
Declaration of Interests.  The authors report no conflict of interest. \\

\vspace{2mm}

\noindent
{\bf \Large Appendices}

\appendix
\section{Derivation of equation (\ref{uinteqn2})}
\label{sec:appA}

In this appendix the volume integral containing the electrostatic or Coulomb force density $\boldsymbol{f}$ in the integral equation (\ref{uinteqn}) for the interfacial 
fluid velocity $\boldsymbol{u}(\boldsymbol{x}_{0})$ is reduced to a surface integral over $S$.  The integral appears in the third or last term on the right hand side 
of equation (\ref{uinteqn}).  In terms of the potential, the force density $\boldsymbol{f}=\varepsilon_{i} \nu \epsilon \nabla^{2}\phi\nabla\phi$ from equations 
(\ref{Stokes}) or (\ref{Stokesa}).  In terms of the local coordinates of the intrinsic frame and the local expression for the potential $\Phi$,
\begin{equation}
\boldsymbol{f}=\varepsilon_{i} \nu \left( \epsilon^{-1}\partial_{N}^{2}\Phi + (\kappa_{1}+\kappa_{2})\partial_{N}\Phi + O(\epsilon) \right) 
\left(\epsilon^{-1}\partial_{N}\Phi \, \boldsymbol{n}  + \nabla_{s} \Phi +O(\epsilon) \right)  .
\label{1}
\end{equation}
The volume element $dV=\epsilon l_{1}l_{2} d\xi_{1}d\xi_{2}dN=\epsilon dS dN (1+\epsilon N(\kappa_{1}+\kappa_{2}) +O(\epsilon^{2}) )$, where 
$dS=a_{1}a_{2}d\xi_{1}d\xi_{2}$ is the surface element on $S$.  If $\boldsymbol{x}_{v}$ is a point in the domain of integration, i.e., in the $\epsilon$-support 
of $\boldsymbol{f}$ near $S$, and $\boldsymbol{x}$ is its normal projection onto $S$
then $\boldsymbol{x}_{v}= \boldsymbol{x}+\epsilon N \boldsymbol{n}$, so that the Stokeslet $\boldsymbol{G}(\boldsymbol{x}_{v}, \boldsymbol{x}_{0})$ in the integrand has the local expansion 
\begin{equation}
\boldsymbol{G}(\boldsymbol{x}_{v}, \boldsymbol{x}_{0})=\boldsymbol{G}(\boldsymbol{x}, \boldsymbol{x}_{0}) + 
\epsilon N \boldsymbol{n}\cdot\nabla_{\boldsymbol{x}}\boldsymbol{G}(\boldsymbol{x}, \boldsymbol{x}_{0}) + O(\epsilon^{2})\, . 
\label{2}
\end{equation}

Introducing the local expansion for $\Phi=\Phi_{0}+\epsilon\Phi_{1} + O(\epsilon^{2})$ of \S \ref{sec:intrinsic}, we find that in the last term on the right hand side 
of equation (\ref{uinteqn}) 
\begin{eqnarray}
\hspace{-2mm}
\boldsymbol{G}(\boldsymbol{x}_{v}, \boldsymbol{x}_{0})\cdot\boldsymbol{f} \, dV_{\boldsymbol{x}_{v}} & = &
 \varepsilon_{i}\nu\boldsymbol{G}(\boldsymbol{x}, \boldsymbol{x}_{0}) \left\{ \epsilon^{-1} \partial_{N}^{2}\Phi_{0}\partial_{N}\Phi_{0}  + 
                      \partial_{N}(\partial_{N}\Phi_{0}\partial_{N}\Phi_{1})     \hspace{23mm} \right.   
\label{3} \\
      & + & \, \left. (\kappa_{1}+\kappa_{2})((\partial_{N}\Phi_{0})^{2}+N \partial_{N}^{2}\Phi_{0}\partial_{N}\Phi_{0}) + O(\epsilon) \right\}
\cdot \boldsymbol{n}(\boldsymbol{x}) \, dN dS_{\boldsymbol{x}} 
\label{4} \\
& + & \varepsilon_{i}\nu \left\{  \boldsymbol{n}(\boldsymbol{x})\cdot\nabla_{\boldsymbol{x}}\boldsymbol{G}(\boldsymbol{x}, \boldsymbol{x}_{0})  
         N\partial_{N}^{2}\Phi_{0}\partial_{N}\Phi_{0} + O(\epsilon) \right\} \cdot \boldsymbol{n}(\boldsymbol{x}) \, dN dS_{\boldsymbol{x}}  
\label{5} \\
& + & \varepsilon_{i}\nu \boldsymbol{G}(\boldsymbol{x}, \boldsymbol{x}_{0})\cdot \partial_{N}^{2}\Phi_{0} \nabla_{s}\Phi_{0} \, dN dS_{\boldsymbol{x}}  + O(\epsilon) \, .
\label{6}
\end{eqnarray}
The dependence on $N$ in the right hand side of this expression is contained in the terms $\Phi_{0}$ and $\Phi_{1}$ of the potential and in the piecewise constant 
permittivity ratio $\varepsilon_{i}$.  The integration over $N\in(-\infty, 0^{-})\cup(0^{+}, \infty)$ can be evaluated in closed form to produce a pair of surface integrals 
over $S$ as described below.  

First, we recall that the hydrodynamic contribution to the traction on $S$ appears in the first integral on the right hand side of (\ref{uinteqn}), which was written in 
terms of the capillary and electrostatic stress at (\ref{hydrofinal}), namely  
\begin{equation}
[\boldsymbol{T}_{H}\cdot\boldsymbol{n}] =  \Delta (\kappa_{1}+\kappa_{2})\boldsymbol{n} -  
\frac{\nu}{2\epsilon} \, [\, \varepsilon_{i} (\partial_{N}\Phi_{0})^{2}\,]  \boldsymbol{n}  
 -\nu \, [\, \varepsilon_{i} \partial_{N}\Phi_{0}\partial_{N}\Phi_{1}] \,  \boldsymbol{n} + O(\epsilon) \, .  
\label{7}
\end{equation}

Integration of (\ref{3}) to (\ref{6}) across the Debye layers requires evaluation of integrals for five distinct $N$-dependent quantities.  Proceeding in the order in 
which these appear, since $\partial_{N}^{2}\Phi_{0}\partial_{N}\Phi_{0}=\partial_{N}(\partial_{N}\Phi_{0})^{2}/2$ and $\partial_{N}\Phi_{0}$ tends to zero exponentially 
as $N\rightarrow \pm\infty$, as implied by (\ref{GC}), we find for the first evaluation that 
\begin{equation}
(\nu/\epsilon)\int_{-\infty}^{\infty} \varepsilon_{i} \partial_{N}^{2}\Phi_{0}\partial_{N}\Phi_{0} \, dN = 
-  \frac{\nu}{2\epsilon} \, [\, \varepsilon_{i} (\partial_{N}\Phi_{0})^{2}\,] \, .
\label{8}
\end{equation}
The corresponding contribution from the body force therefore cancels with the second term on the right hand side of (\ref{7}) for the surface traction on 
recalling the opposite signs that precede the first and third integrals on the right hand side of (\ref{uinteqn}) that are being developed here.  
Similarly, the exponential decay of $\partial_{N}\Phi_{0}$ and matching condition that $\partial_{N}\Phi_{1}\sim \partial_{n}\phi_{0}|_{S_{\pm}}$ is 
bounded at infinity, implies that 
\begin{equation}
\nu \int_{-\infty}^{\infty} \varepsilon_{i} \partial_{N}(\partial_{N}\Phi_{0}\partial_{N}\Phi_{1}) dN = 
 -\nu \, [\, \varepsilon_{i} \partial_{N}\Phi_{0}\partial_{N}\Phi_{1}] \, ,
\label{9}
\end{equation}
which cancels with the third term on the right hand side of (\ref{7}) for the surface traction.  

Integration of the next quantity, which appears in the first term of (\ref{4}), leads us to define  
\begin{equation}
\Theta \equiv \int_{-\infty}^{\infty} \varepsilon_{i} \frac{(\partial_{N}\Phi_{0})^{2}}{2} \, dN = 
     \int_{0^{+}}^{\infty} \varepsilon_{2}\frac{(\partial_{N}\Phi_{0})^{2}}{2} \, dN +  \int_{-\infty}^{0^{-}} \varepsilon_{1}\frac{(\partial_{N}\Phi_{0})^{2}}{2}  \, dN \, .
\label{10}
\end{equation}
It is explained in the text at the end of \S {\ref{sec:forces_drop}} that, up to nondimensionalization and $\epsilon$-rescaling, this is the electrostatic energy 
per unit area contained within the Debye layers, at leading order.  To evaluate 
the integral, we note that equation (\ref{dNPhi0}) gives $\partial_{N}\Phi_{0}$ in terms of the excess potential $\eta_{0}$ and that the definition (\ref{C0ep}) of $\eta_{0}$ 
implies that $\partial_{N}\Phi_{0}=\partial_{N}\eta_{0}$.  A change of variable from $N$ to $\eta_{0}$ shows that 
$\int(\partial_{N}\Phi_{0})^{2}dN=\int\partial_{N}\eta_{0} \, d\eta_{0}$, so that $\Theta$ is given in terms of the layer $\zeta$-potentials by 
\begin{equation}
\Theta= 4 \left( \surd({\varepsilon_{2} c_{0}|_{S_{+}}}) \sinh^{2}\frac{\zeta_{0+}}{4} +   \surd({\varepsilon_{1} c_{0}|_{S_{-}}})\sinh^{2}\frac{\zeta_{0-}}{4} \right) \, .
\label{11}
\end{equation}

The integral that is introduced by both the second term of (\ref{4}) and the Stokes dipole term of (\ref{5}) is evaluated by an integration by parts, namely 
\begin{equation} 
\int_{-\infty}^{\infty} N \partial_{N}^{2}\Phi_{0}\partial_{N}\Phi_{0} \, dN = \left[  N \frac{(\partial_{N}\Phi_{0})^{2}}{2} \right]_{-\infty,0^{+}}^{0^{-},\infty} - \int_{-\infty}^{\infty} \frac{(\partial_{N}\Phi_{0})^{2}}{2} \, dN .
\label{12}
\end{equation} 
The contributions from the boundary terms in the limits $N\rightarrow \pm \infty$ and $N\rightarrow 0^{\pm}$ are zero, so that the integral across the Debye layers 
can be expressed in terms of the quantity $\Theta$ of (\ref{10}) as 
\begin{equation}
 \int_{-\infty}^{\infty} \varepsilon_{i} N \partial_{N}^{2}\Phi_{0}\partial_{N}\Phi_{0} \, dN =- \Theta \, .
\label{13}
\end{equation}

The integral that is introduced by the last, tangential derivative term of (\ref{6}) is found by an integration by parts to be 
\begin{equation}
\int_{-\infty}^{\infty} \varepsilon_{i} \partial_{N}^{2}\Phi_{0} \nabla_{s}\Phi_{0} \, dN = 
\left[ \varepsilon_{i} \partial_{N}\Phi_{0}  \nabla_{s}\Phi_{0}   \right]_{-\infty,0^{+}}^{0^{-},\infty} 
            -  \nabla_{s} \int_{-\infty}^{\infty} \varepsilon_{i} \frac{(\partial_{N}\Phi_{0})^{2}}{2} \, dN \, .
\label{14}
\end{equation}
In the boundary term, the contributions as $N\rightarrow {\pm\infty}$ are zero because of the exponential approach of $\Phi_{0}$ to the outer potential $\phi_{0}|_{S_{\pm}}$. 
The contributions as $N\rightarrow 0^{\pm}$ are also zero.  This occurs: (i) because of a cancellation, for both a vesicle and a drop, with the exact jump discontinuity 
$[\varepsilon_{i} \partial_{n}\phi\nabla_{s}\phi]$ across $S$ of (\ref{hydrostressS}), when the exact relation is evaluated at leading order, and (ii) a fortiori for a drop, because, 
as noted below equation (\ref{hydrostressS}), the jump discontinuity itself vanishes for the sharp interface of a drop that carries no monopole surface charge, since this 
implies that $[\varepsilon_{i}\partial_{n}\phi]=0$ per the second of equations (\ref{BC1a}), and both $\phi$ and $\nabla_{s}\phi$ are continuous across $S$.  Hence, 
\begin{equation}
\int_{-\infty}^{\infty} \varepsilon_{i} \partial_{N}^{2}\Phi_{0} \nabla_{s}\Phi_{0} \, dN = - \nabla_{s} \Theta \, . 
\label{15}
\end{equation}

Apart from the Stokes dipole term of (\ref{5}), integration across the Debye layers of the third integral of (\ref{uinteqn}), i.e., that containing the electrostatic force 
$\boldsymbol{f}$, is seen to produce four contributions to a surface integral over $S$ that has the same form as the first, Stokeslet integral of (\ref{uinteqn}) containing 
the hydrodynamic stress.  These are combined with the one remaining capillary stress term on the right hand side of (\ref{7}) to define a modified surface traction due to the 
combined effects of the hydrodynamic stress and the electrostatic force exerted by the adjacent Debye layers on $S$.  This is given by replacing the surface traction 
$[\boldsymbol{T}_{H}\cdot\boldsymbol{n}]$ in (\ref{uinteqn}) with 
\begin{equation}
[\boldsymbol{\hat{T}}\cdot\boldsymbol{n}]=(\kappa_{1}+\kappa_{2})\left( \Delta - \nu\Theta \right)\boldsymbol{n} + \nu \nabla_{s}\Theta \, ,
\label{16}
\end{equation}
where $\Theta$ is given by (\ref{11}).  The surface integral that contains the Stokes dipole replaces the third and last integral of (\ref{uinteqn}) with 
\begin{equation}
\frac{-1}{4\pi(\mu_{2}+\mu_{1})}\int_{S} n_{k}(\boldsymbol{x})\frac{\partial}{\partial x_{k}} G_{ij}(\boldsymbol{x}, \boldsymbol{x}_{0}) 
    \, \nu\Theta(\boldsymbol{x})\, n_{i}(\boldsymbol{x})\, dS_{\boldsymbol{x}} \, .
\label{17}
\end{equation}
This results in the integral equation (\ref{uinteqn2}).

\section{Reduced form of the stress-balance boundary condition}
\label{sec:appB}

For some purposes, such as construction of a small-amplitude expansion about an equilibrium state, the stress-balance boundary condition needs to be used in a 
reduced form, as opposed to being embedded in an integral equation.  We use the term `reduced' in the same sense in which it has been used elsewhere in this study.  
Here, the Debye layer structure is used to express the stress-balance boundary condition on $S$ in terms of dependent variables at the outer edges of the layers and 
known surface data.  The formulation uses the intrinsic orthogonal curvilinear coordinate system of \S \ref{sec:intrinsic}.  

The boundary condition can be written as 
\begin{equation}
[(\boldsymbol{T}_{H} + \nu \epsilon \boldsymbol{T}_{M})\cdot\boldsymbol{n}] = \boldsymbol{\tau} \, ,
\label{stressbalplus}
\end{equation}
on $S$, where the stress in the interface is 
\begin{equation}
\boldsymbol{\tau} =  \left\{  
\begin{array}{ll}
\boldsymbol{\tau}_{d} \equiv \Delta(\kappa_{1}+\kappa_{2})\, \boldsymbol{n} & \mbox{for a drop} \, , \\
\boldsymbol{\tau}_{m} \equiv  2 \sigma H \boldsymbol{n} - \nabla_{s} \sigma  &  \\
\hspace{7mm} -  \kappa_{b} \left( 2H(H^{2}-K)  +  \nabla_{s}^{2} H \right) \boldsymbol{n} & \mbox{for a vesicle} \, .
\end{array}       \right.
\label{hoo}
\end{equation}

The hydrodynamic stress is given in coordinate-free form by the first of relations (\ref{tensnondim}).  When expressed in the intrinsic frame and then evaluated 
on the interface $n=0$ the traction is given by 
\begin{eqnarray}
\boldsymbol{T}_{H}\cdot\boldsymbol{n} &=& \mu \left( \nabla_{s}u_{p} +\boldsymbol{e}_{1} (\partial_{n}-\kappa_{1})u_{t1} 
                             +\boldsymbol{e}_{2} (\partial_{n}-\kappa_{2})u_{t2}  \right) \nonumber \\
                             & & + \, ( \, - p + 2\mu\partial_{n}u_{p} ) \, \boldsymbol{n} \nonumber \\ 
   &=&  \mu \left( \nabla_{s}u_{n0} +\boldsymbol{e}_{1} (\partial_{N}U_{t11}-\kappa_{1}u_{s10})  
                             +\boldsymbol{e}_{2} (\partial_{N}U_{t21}-\kappa_{2}u_{s20})  \right) \nonumber \\
                             & & + \, ( \, - \frac{P_{-1}}{\epsilon} - P_{0} + 2\mu\partial_{n}v_{p}|_{S} ) \, \boldsymbol{n} + O(\epsilon) \, , 
\label{boo}
\end{eqnarray}
where the first relation is exact and the second relation is found by expansion of variables within the Debye layers.  The electrostatic stress is given in 
coordinate-free form by the second of relations  (\ref{tensnondim}), and its expression in the intrinsic frame on $n=0$ was developed at equations 
(\ref{hydrostressS}) and (\ref{hydrofinal}).  The interfacial traction is given by 
\begin{eqnarray}
\nu\epsilon \, \boldsymbol{T}_{M} \cdot \boldsymbol{n} &=& \nu\epsilon \, \varepsilon \left( \frac{1}{2}(\partial_{n}\phi^{2}-| \nabla_{s}\phi |^{2}) \, \boldsymbol{n} + 
\partial_{n}\phi\nabla_{s}\phi \right) \nonumber \\ 
                             &=& \nu \varepsilon \left( \frac{(\partial_{N}\Phi_{0})^{2}}{2\epsilon} + \partial_{N}\Phi_{0} \partial_{N}\Phi_{1} \right) \boldsymbol{n} 
                                     + \nu \varepsilon \partial_{N}\Phi_{0} \nabla_{s}\Phi_{0} + O(\epsilon) \, ,
\label{moo}
\end{eqnarray}
where the terms in $\nabla_{s}\phi$ that vanish in the jump condition for a drop have been kept. 

{\it \underline{The Normal Component.}}  When the jump operator is applied, the $O(\epsilon^{-1})$ terms of the tractions are found to cancel, via equation (\ref{P-1}), 
and when the expression (\ref{P0}) for $P_{0}$ is used integrals appear that give the quantity $\Theta$ defined at (\ref{10}); see also (\ref{11}).  The normal component 
of the stress-balance boundary condition is 
\begin{equation}
2(\kappa_{1}+\kappa_{2})\nu\Theta - [p_{0}]^{S_{+}}_{S_{-}} + 2 (\mu_{2}-\mu_{1}) \partial_{n}v_{p}|_{S} = \boldsymbol{\tau} \cdot \boldsymbol{n} \, , 
\label{norm1}
\end{equation}
where the interface stress $\boldsymbol{\tau}$ for a drop or vesicle is given by (\ref{hoo}).  

{\it \underline{The Tangential Component.}} First, we form the shear stress term $\mu\partial_{N}U_{t11}$ by differentiating equation (\ref{St4}) to find that, since 
$\partial_{N}\eta_{0}=\partial_{N}\Phi_{0}$, 
\begin{equation}
\mu\partial_{N}U_{t11} = - \nu \varepsilon \partial_{N}\Phi_{0} \left( \frac{1}{a_{1}} \partial_{\xi_{1}}  \! \phi_{0}|_{S_{\pm}} 
             - \tanh \frac{\eta_{0}}{4} \, \frac{1}{a_{1}} \partial_{\xi_{1}} \! \! \ln c_{0}|_{S_{\pm}} \right) + \mu\partial_{n} u_{10}|_{S_{\pm}}
\label{tanga}
\end{equation}
in the Debye layers.  A similar expression holds for $\mu\partial_{N}U_{t21}$.  Note that a factor $\varepsilon \partial_{N}\Phi_{0}$ premultiplies the $N$-dependent 
contribution from within the layers, and when the jump operator is applied across the interface at $n=0$ this factor is continuous, per the jump conditions of no monopole 
charge at (\ref{BC1a}) for a drop and (\ref{capbc}) for a vesicle.  We choose to express the factor in the limit that the interface is approached from the exterior 
$\Omega_{2}$, or on $\partial\Omega_{2}$, although the choice is arbitrary, and evaluation is found from (\ref{dNPhi0}). 

The electrostatic stress as evaluated at (\ref{moo}) has a tangential component given by $\nu\varepsilon \partial_{N}\Phi_{0} \nabla_{s}\Phi_{0}$.  Here the 
factor $\varepsilon \partial_{N}\Phi_{0}$ that has just been noted as continuous across $S$ appears again, multiplied by $\nabla_{s}\Phi_{0}$.  We noted earlier 
that for a drop, the boundary condition at (\ref{BC1a}) that the potential is continuous, or $[\phi]=0$, implies that $[\nabla_{s}\Phi_{0}]=0$, so that, when the jump is 
formed, there is seen to be no tangential electrostatic stress exerted on the drop's interface.  However, for a vesicle there is a nonzero jump and electrostatic stress 
component that can be expressed as $\nu\varepsilon_{2} \partial_{N}\Phi_{0}|_{\partial\Omega_{2}} [\nabla_{s}\Phi_{0}]$.  This can be combined with the quantity 
$- \nu\varepsilon_{2} \partial_{N}\Phi_{0}|_{\partial\Omega_{2}} [\nabla_{s}\phi_{0}]^{S_{+}}_{S_{-}}$, which is implied by (\ref{tanga}) and its companion 
$\mu\partial_{N}U_{t21}$, to give a term for a vesicle that is proportional to the jump $[\nabla_{s}\zeta_{0}]^{S_{+}}_{S_{-}}$ in the $\zeta$-potentials.  For a drop, 
however, the analogous term is proportional to the jump $[\nabla_{s}\phi_{0}]^{S_{+}}_{S_{-}}$ in the outer potentials with a change of sign. 

Piecing these results together with the interface stress $\tau$ given by (\ref{boo}), we have the tangential component of the stress-balance boundary condition for 
a drop given by
\begin{eqnarray}
(\mu_{2}-\mu_{1}) \left\{ \nabla_{s}u_{n0} -(\kappa_{1}u_{s10}\boldsymbol{e}_{1} + \kappa_{2}u_{s20}\boldsymbol{e}_{2})\right\} 
    + \left[ \mu ( \partial_{n}u_{10}\boldsymbol{e}_{1} + \partial_{n}u_{20}\boldsymbol{e}_{2} ) \right]^{S_{+}}_{S_{-}} \nonumber \\
 + 2\nu \surd({\varepsilon_{2} c_{0}|_{S_{+}}}) \sinh \frac{\zeta_{0+}}{2} \left[\nabla_{s}\phi_{0} - \tanh \frac{\zeta_{0}}{4} \, \nabla_{s} \! \ln c_{0} \right]^{S_{+}}_{S_{-}} = 0 \, . 
      \hspace{5mm} 
\label{tdrop}
\end{eqnarray}
The analogous expression for a vesicle is 
\begin{eqnarray}
(\mu_{2}-\mu_{1}) \left\{ \nabla_{s}u_{n0} -(\kappa_{1}u_{s10}\boldsymbol{e}_{1} + \kappa_{2}u_{s20}\boldsymbol{e}_{2})\right\} 
    + \left[ \mu ( \partial_{n}u_{10}\boldsymbol{e}_{1} + \partial_{n}u_{20}\boldsymbol{e}_{2} ) \right]^{S_{+}}_{S_{-}} \nonumber \\
 - 2\nu \surd({\varepsilon_{2} c_{0}|_{S_{+}}}) \sinh \frac{\zeta_{0+}}{2} \left[\nabla_{s}\zeta_{0} + \tanh \frac{\zeta_{0}}{4} \, \nabla_{s} \! \ln c_{0} \right]^{S_{+}}_{S_{-}} 
           = -\nabla_{s} \sigma  \, . 
      \hspace{2mm} 
\label{tvesicle}
\end{eqnarray}
  

\begin{thebibliography}{99}

\bibitem{Helmholtz1853} H.~Helmholtz.  Ueber einige gesetze der vertheilung elektrischer Str\"{o}me in 
k\"{o}rperlichen leitern mit anwendung auf die thierisch-elektrischen versuche. {\em Annalen der Physik und Chemie}, 165: 211-233, 1853.

\bibitem{Gouy} G.~Gouy.  Sur la constitution de la charge \'electrique \`a la surface d'un \'electrolyte. {\em J.~Phys.~Radium}, 9: 457-468, 1910.

\bibitem{Chapman}  D.~L.~Chapman.  A contribution to the theory of eltroencapillarity.  {\em Phil.~Mag}, 25: 475-481, 1913.

\bibitem{Russel_etal}  W.~B.~Russel, D.~A.~Saville, and W.~R.~Schowalter.  {\em Colloidal Dispersions}. Cambridge University Press, Cambridge, 1989.

\bibitem{Hunter}  R.~J.~Hunter.  {\em Foundations of Colloid Science (Second Edition)}, Oxford University Press, Oxford, 2001.

\bibitem{Squires_Bazant_04}  T.~M.~Squires and M.~Z.~Bazant.  Induced-charge electro-osmosis.  {\em J.~Fluid~Mech.}, 509: 217-252, 2004. 

\bibitem{Bazant_Squires_10}  M.~Z.~Bazant and T.~M.~Squires. Induced-charge electrokinetic phenomena.  {\em Curr.~Opin.~Colloid~Interface~Sci.}, 15: 203-213, 2010.

\bibitem{Girault_2010}  H.~H.~Girault.  Electrochemistry at liquid-liquid interfaces. In A.~J.~Bard and C~.G.~Zoski, editors, {\em Electroanalytical Chemistry}, 23: 1-104, 
CRC Press, Boca Raton, 2010.

\bibitem{Reymond_&co}  F.~Reymond, D.~Fermin, H.~J.~ Lee, and H.~H.~Girault.  Electrochemistry at liquid/liquid interfaces: methodology and potential applications. 
{\em Electrochimica Acta}, 45: 2647-2662, 2000.

\bibitem{Vlahovska2019}  P.~M.~Vlahovska.  Electrohydrodynamics of drops and vesicles.  {\em Annu. Rev. Fluid Mech.}, 51: 305-330, 2019.

\bibitem{Melcher_Taylor_1969}  J.~R.~Melcher and G.~I.~Taylor.  Electrohydrodynamics: a review of the role of interfacial shear stress. {\em Annu. Rev. Fluid Mech.},
1: 111-146, 1969. 

\bibitem{Saville_1997}  D.~A.~Saville.  Electrohydrodynamics: the {T}aylor-{M}elcher leaky dielectric model.  {\em Annu. Rev. Fluid Mech.}, 29: 27-64, 1997. 

\bibitem{Taylor_1966}  G.~I.~Taylor.  Studies in electrohydrodynamics. {I}. {T}he circulation produced in a drop by an electric field.  {\em Proc.~Roy.~Soc.~Lond.,~A.}, 
291: 159-166, 1966. 

\bibitem{SchnitzerYariv}  O.~Schnitzer and E.~Yariv. The {T}aylor-{M}elcher leaky dielectric model as a macroscale electrokinetic description.  {\em J. Fluid Mech.}, 
773:1-33, 2015. 

\bibitem{MoriYoung}  Y.~Mori and Y.-N.~Young.  From electrodiffusion theory to the electrohydrodynamics of leaky dielectrics through the weak electrolyte limit.  
{\em J.~Fluid Mech.}, 855: 67-130, 2018.

\bibitem{PascallSquires2011}  A.~J.~Pascall and T.~M.~Squires.  Electrokinetics at liquid/liquid interfaces.  {\em J. Fluid Mech.}, 684: 163-191, 2011. 

\bibitem{Zholkovskij}  E.~K.~Zholkovskij, J.~H.~Masliyah, and J.~Czarnecki.  An electrokinetic model of drop deformation in an electric field.  {\em J.~Fluid Mech.}, 
472: 1-27, 2002.  

\bibitem{Mori_Liu_Eisenberg}  Y.~Mori, C.~Liu, and R.~S.~Eisenberg.  A model of electrodiffusion and osmotic water flow and its energetic structure.  {\em Physica D}, 
240: 1385-1852, 2011. 

\bibitem{Vlahovska2009}  P.~M.~Vlahovska, T.~Podgorski, and C.~Misbah.  Vesicles and red blood cells in flow: From individual dynamics to rheology. 
{\em Comptes Rendus Physique}, 10: 775-789, 2009.

\bibitem{Seifert1999}  U.~Seifert.  Fluid membranes in hydrodynamic flow fields: Formalism and an application to fluctuating quasispherical vesicles in shear flow. 
{\em Eur.~Phys.~J.~B}, 8: 405-415, 1999. 

\bibitem{BBiesel2011}  D.~Barth\`{e}s-Biesel.  Modeling the motion of capsules in flow.  {\em Current Opinion in Colloid \& Interface Science}, 16: 3-12, 2011.

\bibitem{Helfrich1973}  W.~Helfrich.  Elastic properties of lipid bilayers - Theory and possible experiments. {\em Z.~Naturforsch.}, 28c: 693-703, 1973. 

\bibitem{Batchelor} G.~K.~Batchelor.  {\em An Introduction to Fluid Dynamics},  Cambridge University Press, Cambridge, 2000. 

\bibitem{Weatherburn}  C.~E.~Weatherburn.  {\em Differential Geometry of Three Dimensions, Volume 1}, Cambridge University Press, Cambridge, 1927 (Reissued 2016).

\bibitem{JCP_bubble}  M. R. Booty and M. Siegel.  A hybrid numerical method for interfacial fluid flow with soluble surfactant.  {\em J.~Comput. Phys.}, 229: 3864-3883, 2010. 

\bibitem{RPA2020}  R.~P.~Atwater.  {\em Studies of two-phase flow with soluble surfactant}, PhD thesis, New Jersey Institute of Technology, 2020. 

\bibitem{Pozrikidis_book1}  C.~Pozrikidis.  {\em Boundary Integral and Singularity Methods for Linearized Viscous Flow}, Cambridge University Press, Cambridge,1992. 

\bibitem{Rallison&Acrivos}  J.~M.~Rallison and A.~Acrivos.  A numerical study of the deformation and burst of a viscous drop in an extensional flow.  {\em J.~Fluid Mech.}, 
89:191-200, 1978. 

\bibitem{Kress}  R.~Kress.  {\em Linear Integral Equations}, Springer, New York, 1999.

\bibitem{Allan_Mason1962}  R.~S.~Allan and S.~G.~Mason.  Particle behaviour in shear and electric fields {I}. {D}eformation and burst of fluid drops. 
{\em Proc.~Roy.~Soc.~Lond.,~A.}, 267: 45-61, 1962. 

\bibitem{Vlahovska_Gracia_2007}  P.~M.~Vlahovska and R.~S.~Gracia.  Dynamics of a viscous vesicle in linear flows.  {\em Phys. Rev. E}, 75: 016313, 2007. 

\bibitem{Schwalbe_etal_2011}  J.~T.~Schwalbe, P.~M.~Vlahovska and M.~J.~Miksis.  Vesicle electrohydrodynamics.  {\em Phys. Rev. E}, 83: 046309, 2011.  

\bibitem{WoodhouseGoldstein2012}  F.~G.~Woodhouse and R.~E.~Goldstein.  Shear-driven circulation patterns in lipid membrane vesicles.  {\em J. Fluid Mech.}, 
705: 165-175, 2012. 



\end{thebibliography}

\end{document}